\documentclass[preprint,12pt]{elsarticle}

\usepackage[T1]{fontenc}
\usepackage{lmodern}  
\usepackage{url}
\usepackage[bottom]{footmisc}
\usepackage{amsmath}
\usepackage{amssymb}
\usepackage{amsfonts}
\usepackage{breakcites}
\usepackage{newtxmath}
\usepackage{natbib}
\usepackage{hyperref}
\usepackage{comment}
\usepackage{color}
\usepackage{soul} 
\usepackage{siunitx}

\graphicspath{{./}}

\journal{Computers and Fluids}

\begin{document}
\begin{frontmatter}


\title{A Hybrid Immersed-Boundary/Front-Tracking Method For Interface-Resolved Simulation of Droplet Evaporation}

\author[label1]{Faraz Salimnezhad}
\author[label2]{Hasret Turkeri}
\author[label3]{Iskender Gokalp}
\author[label1]{Metin Muradoglu}
\affiliation[label1]{organization={Koc University},
            addressline={Rumelifeneri Yolu, Sariyer}, 
            city={Istanbul},
            postcode={34450}, 
            country={Turkey}}
\affiliation[label2]{organization={Tusas Engine Industries (TEI)},
            addressline={Tepebasi}, 
            city={Eskisehir},
            postcode={26210}, 
            country={Turkey}}
\affiliation[label3]{organization={Tubitak Marmara Research Center, Gebze},
            city={Kocaeli},
            postcode={41470}, 
            country={Turkey}}
\begin{abstract}
A hybrid sharp-interface immersed-boundary/front-tracking (IB/FT) method is developed for interface-resolved simulation of evaporating droplets in incompressible multiphase flows. A one-field formulation is used to solve the flow, species mass fraction and energy equations in the entire computational domain with appropriate jump conditions at the interface.
An image point and ghost cell methodology is coupled with a front-tracking method to achieve an overall second order spatial accuracy for the mass fraction boundary condition on the droplet surface. The immersed-boundary method is also extended to simulate mass transfer from a solid sphere in a convective environment. The numerical method is first validated for the standard benchmark cases and the results are found to be in good agreement with analytical solutions. The method is shown to be overall second order accurate in space. Employing a moving reference frame methodology, the method is then applied to simulate evaporation of a deformable droplet in a convective environment and the results are compared with the existing evaporation models widely used in spray combustion simulations. 
\end{abstract}

\begin{graphicalabstract}
\begin{figure}[ht]
\centering
\begin{tabular}{cc}
\includegraphics[scale=0.53]{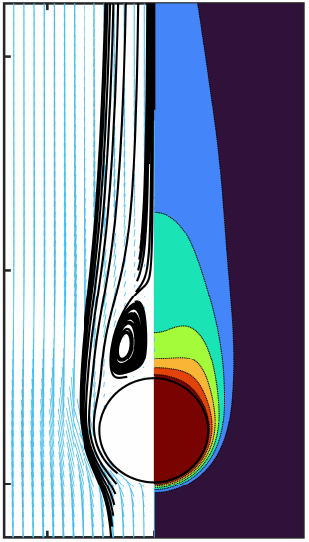} 
&

\includegraphics[scale=0.53]{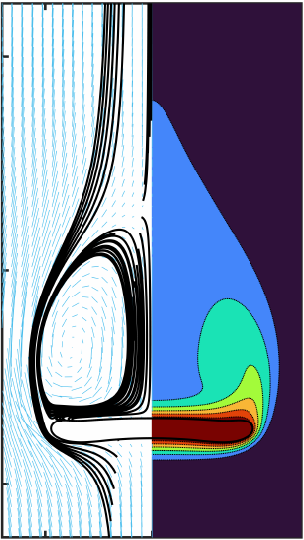} \\
\includegraphics[scale=0.53]{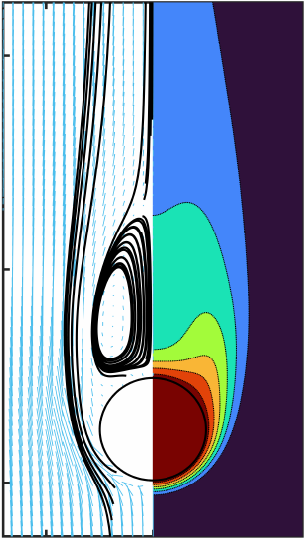} 
&

\includegraphics[scale=0.53]{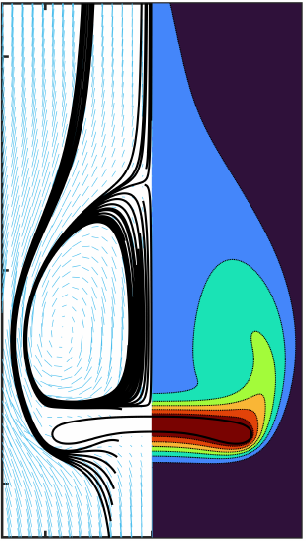} \\
\end{tabular}
\caption{Effect of Stefan flow on the vapor mass fraction and flow fields around evaporating droplets including both nearly spherical and deforming cases.} 
\label{Fig:We06GA}
\end{figure}
\end{graphicalabstract}

\begin{highlights}
\item A hybrid IB/FT method is developed for simulating evaporation in multiphase flows.
\item An image-point and ghost-cell method is used to enforce mass boundary conditions on droplets.
\item The method is validated against benchmark test cases, demonstrating second-order spatial accuracy.
\item The method is extended to simulate mass transfer from solid spheres in convective flows.
\item The numerical results are compared with the existing evaporation models.
\end{highlights}

\begin{keyword}
Multiphase flows \sep droplet evaporation \sep  phase change \sep direct numerical simulation \sep front-tracking method \sep sharp-interface immersed boundary method \sep evaporation models.


\end{keyword}

\end{frontmatter}


\section{Introduction}
Droplet evaporation is of fundamental importance in a wide range of natural phenomena and industrial applications \cite{PhysRevFluids.6.020501,mittal_ni_seo_2020,sazhin2006advanced,BIROUK2006408}. In particular, performance of spray combustion used in energy conversion devices such as internal combustion engines, gas turbines, liquid-fueled rocket engines and liquid-fueled industrial burners critically depends on liquid fuel atomization and droplet evaporation process in a convective environment \cite{JENNY2012846,sazhin2006advanced,BIROUK2006408}. Therefore, it is highly desirable to understand fuel droplet evaporation in a flowing ambient fluid for improving efficiency and power output of such devices. Droplet evaporation also plays a singularly vital role in the eventual fate of virus-laden droplets that are formed in the respiratory track \cite{doi:10.1126/science.abd9149,mittal_ni_seo_2020,wells1934air}.

Drop dynamics is highly complicated mainly due to the existence of the phase boundary (interface) that continuously deforms and may undergo topological changes in a complex flow field. Evaporation adds a further complexity making simulation of an evaporating droplet a challenging task especially in a convection dominated flow where thin mass boundary layer in the leading edge and a large flow separation behind the droplet need to be fully resolved \cite{JENNY2012846,sazhin2006advanced,BIROUK2006408}. 

Interface-resolved simulations of multiphase systems can give indispensable insight into underlying flow physics and help enhance understanding of evaporation process in applications ranging from heat exchangers and combustion engines to environmental processes such as cloud formation \cite{environmentalfluid}. Several high-fidelity interface-resolved direct numerical simulation methods have been developed to tackle these challenges. The methods can be broadly categorized as the interface-capturing and the interface-tracking approaches~\cite{tryggvason2011direct}. In the interface-capturing methods, such as level-set (LS) and volume-of-fluid (VOF), the interface is represented implicitly by a color function. Interface-capturing methods have been widely used for interface-resolved simulations of interfacial flows mainly due to their implicit representation of interfaces in a fully Eulerian framework, which makes handling of topology change relatively easy and parallelization highly scalable, and good conservation properties especially in the presence of large jumps in material properties across phase boundaries. However, they suffer from the lack of precise information about location of interface which makes accurate computation of mass and energy exchange at the interface very difficult.  \citet{10.1115/1.2830042} pioneered the application of the level-set method to a phage change problem and studied film boiling in a two-dimensional setting. The level-set method has become a popular choice in phase change simulations due to relatively easy treatment of interfacial jump conditions via coupling with the ghost fluid method \cite{GibouFedkiw-jcp-2005,TanguyEtAl-jcp-2014,VillegasEtAl-jcp-2016,LeeEtAl-jcp-2017,GeEtAl-jcp-2018,LuoEtAl-PCES-2019}. An important drawback of the level-set method has been reported as its poor conservation properties caused by the non-conservative nature of the level-set function \cite{LuoEtAl-PCES-2019}. On the other hand, the volume-of-fluid method, another popular interface-capturing technique, is appealing particularly due to its potential to achieve exact mass conservation regardless of grid resolution. However, in this approach, advection of the VOF color function has proved to be a challenging task limiting its widespread use in practical  applications. Nevertheless, the VOF method has been successfully used to simulate various phase change problems \cite{SCHLOTTKE20085215,MaBothe-jcp-2013,PALMORE2019108954,ReutzschEtAl-jcp-2020,ScapinEtAl-jcp-2020,MalanEtAl-jcp-2021,BoniouEtAl-jcp-2022,BoydLing-cf-2023} since the pioneering work by \citet{WELCH2000662} who developed the first VOF to investigate boiling flows in two-dimensional setup.

The front-tracking method used here has a distinct advantage of keeping  interface sharp without any significant numerical diffusion and facilitating accurate evaluation of the flow and transport properties at the interface, thanks to the explicit representation of interface using a separate Lagrangian grid \cite{tryggvason2001front,tryggvason2011direct}. The method has been successfully used to simulate a vast range of laminar and turbulent multiphase flows involving various multi-physics effects including phase change \cite{juric1998computations}, soluble surfactant \cite{muradoglu2008front,muradoglu2014simulations}, electric field \cite{ALRAWAHI2002471}, and complex fluid rheology \cite{IZBASSAROV2015122}. Regarding the applications to the phase change problems, \citet{juric1998computations} developed a front-tracking method to simulate film boiling using an iterative method to satisfy the temperature boundary condition at the interface. Esmaeeli and Tryggvason \cite{esmaeeli2003computations,ESMAEELI20045451,ESMAEELI20045463} used essentially the same method but set the saturation temperature as a boundary condition on the interface to eliminate the need for an iterative procedure. \citet{koynov2005mass} employed a front-tracking method to study the buoyancy-driven motion of deformable bubbles accounting for one-way mass transfer from the bubble to the ambient fluid and chemical reactions in a wide range of operating conditions but they neglected the volume change due to mass transfer. \citet{ABOULHASANZADEH2012456} devised a multiscale method within the front-tracking framework for simulation of mass transfer from buoyant bubbles using a boundary layer approximation in the vicinity of the bubble to address the need for high grid resolution at high Schmidt numbers. \citet{irfan2017front} developed a front-tracking method for interface-resolved simulation of evaporation process in a two dimensional setup using both the species-driven and the temperature-driven evaporation models. In the follow up study, they also included chemical reactions to simulate a fuel droplet evaporation and combustion in an axisymmetric setup \cite{irfan2018front}. \citet{khorram2022direct}  used a similar algorithm to simulate film boiling and bubble growth in a three dimensional setup but they also accounted for bubble breakup and coalescence using a novel topology change algorithm. Recently, \citet{NAJAFIAN20231} improved this method to simulate droplet evaporation at high density ratio by advancing the front in two steps and solving the associated pressure Poisson equation twice to impose the mass conservation constraint. 

In all the previous front-tracking methods, the species boundary condition at the interface is treated indirectly by conservatively distributing the evaporative mass flux as a source term outside the droplet near the interface using a one-sided distribution algorithm \cite{irfan2017front,irfan2018front,NAJAFIAN20231}, which reduces the spatial accuracy significantly and results in a first order accuracy locally in general and globally in the strongly evaporating regimes \cite{irfan2017front}. As a result, an extreme grid resolution is usually required to reduce the spatial error below an acceptable level in strongly evaporating and convection dominating regimes, which is typically the case in combustion applications. This deficiency constitutes the main motivation in the present study. To address this issue, we develop a hybrid sharp-interface immersed-boundary/front-tracking (IB/FT) method that maintains an overall second order spatial accuracy and keeps the interface sharp. The main novelty is that the mass fraction is accurately set as a Dirichlet boundary condition on the droplet surface using the image-point and ghost-cell methodology developed by \citet{mittal2008versatile}. It is shown that the current method achieves a locally and globally second order spatial accuracy while maintaining a sharp representation of the interface. The method is validated against the well known $d^2$-law for which the mass fraction value, corresponding to a specific mass transfer number ($B_M$),  is applied as a Dirichlet boundary at the interface. It is also tested for the wet-bulb temperature of a water droplet for a range of ambient conditions and the results are shown to be in good agreement with the psychometric chart. After that, employing a moving reference frame methodology, the method is applied to study droplet evaporation in a convective environment and the results are compared with the Abramzon-Sirignano (A-S) and the classical models for nearly spherical, moderately deforming and extremely deforming droplet cases. The immersed boundary methodology \citep{mittal2008versatile} is also extended to simulate heat and mass transfer from a solid object immersed in a flowing fluid, and the results are found to be in good agreement with the results reported in the literature.  

The rest of the paper is organized as follows. The mathematical formulation of the current study is given in Section~\ref{Mathform}. Numerical method, including the use of image-point and ghost-cell methodology and the novel front restructuring algorithm, are described in Section~\ref{NumMeth}.  The results are presented and discussed in Section~\ref{results}. Concluding remarks are made in Section~\ref{conlusions}.
\label{Int}
\section{Mathematical Formulation}
\label{Mathform}
In the front-tracking method, the governing field equations are solved in the entire computational domain using a one-field formulation. The equations can be solved in both conservative and non-conservative forms. In the present study, the non-conservative form is preferred to reduce the inconsistency between the advection schemes used in advancing the interface (thus the material properties) and approximating the convective terms in the momentum equations \cite{tryggvason2011direct}. In this framework, the incompressible Navier-Stokes equations can be written as \citep{sato2013sharp,BHUVANKAR2020115919}:
\begin{equation} 
\begin{aligned}
\rho \frac{\partial { \boldsymbol{u}}}{\partial {t}} +\rho \left[\nabla \cdot  \left(\boldsymbol{uu}\right) - \boldsymbol{u} \left(\nabla \cdot \boldsymbol{u}\right)\right]=  &- \nabla p+ \rho \boldsymbol{g} +  \nabla \cdot \mu \left(\nabla \boldsymbol{u}+\nabla {\boldsymbol{u}}^T\right)+ \\ &+\int_A{\sigma \kappa \boldsymbol{n}\delta \left(\boldsymbol{x}\mathrm{-}{\boldsymbol{x}}_{\mathit{\Gamma}}\right)dA},
\end{aligned}
\label{eqn:momentum}
\end{equation}
where $\boldsymbol{u}$, $p$ and $\boldsymbol{g}$ denote the velocity vector, the pressure field and the gravitational acceleration, respectively. The density, $\rho$, and viscosity $\mu$ fields vary discontinuously across the phase boundary. The final term on the right-hand side of Eq.\eqref{eqn:momentum} represents the body force due to surface tension where $\sigma $ is the surface tension coefficient, $\kappa $ is twice the mean curvature and $\boldsymbol{n}$ is the outward normal vector at the interface. Note that the term, $\rho\boldsymbol{u} \left(\nabla \cdot \boldsymbol{u}\right)$, on the left-hand side of Eq.\eqref{eqn:momentum} is needed to account for the surface regression caused by the phase change. 

The flow is assumed to be incompressible in both phases. Nevertheless, the mass conservation equation is modified near the interface to account for the volume expansion due to the phase change as \citep{esmaeeli2004front,irfan2017front,irfan2018front}:
\begin{equation} 
\nabla \cdot \boldsymbol{u}=\frac{1}{h_{lg}}\left(\frac{1}{{\rho }_g}-\frac{{1}}{{\rho }_l}\right)\int_A{\delta \left(\boldsymbol{x}-{\boldsymbol{x}}_{\mathit{\Gamma}}\right)\dot{q}_{\Gamma} \ dA}, 
\label{eqn:continuity}
\end{equation}
where the delta function indicates that the source term is non-zero at the interface and zero elsewhere, $h_{lg}$ and $\dot{q}_{\Gamma}$ represent the latent heat of vaporization and the heat flux per unit time at the interface, respectively. The subscripts $l,g,$ and $\Gamma$, respectively, indicate liquid (droplet) phase, gas phase (ambient fluid) and interface. 
Taking the evaporation into account, the energy and vapor mass-fraction evolution equations are given by
\begin{eqnarray} 
\rho c_p \left(\frac{\partial T}{\partial t}+ \boldsymbol{u} \cdot \nabla T\right) & = & \nabla \cdot k \nabla T- \int_A{\delta \left(\boldsymbol{x}-{\boldsymbol{x}}_{\Gamma}\right)\dot{q}_{\Gamma} \ dA},
\label{eqn:energyeq} \\
\frac{\partial Y}{\partial t}+\boldsymbol{u} \cdot \nabla Y & = & \nabla \cdot D_{vg}\nabla Y,  
\label{eqn:specieseq}
\end{eqnarray}
where $T$, $c_p$, $k$, $Y$ and  $D_{vg}$ are the temperature, the specific heat capacity, the thermal conductivity, the vapor mass-fraction and the vapor mass diffusion coefficient, respectively. The last term on the right hand of Eq.~(\ref{eqn:energyeq}) is responsible for the cooling effect due to the evaporation. The heat flux term in Eqs.~(\ref{eqn:energyeq}) and (\ref{eqn:continuity}) is related to the mass flux as $\dot{q}_{\Gamma}=\dot{m}_{\Gamma} h_{lg}$. The mass conservation across the interface requires
\begin{equation} 
\begin{aligned}
{\dot{m}}_{\Gamma}Y^{\Gamma}_l-{\dot{m}}_{\Gamma}Y^{\Gamma}_g-{\rho }_gD_{vg}{\left(\frac{\partial Y}{\partial n}\right)}_{\Gamma}=0.
\end{aligned}
\label{eqn:specieseqjump}
\end{equation}  
Since $Y^{\Gamma}_l =1$ for a single-component droplet, Eq.~\eqref{eqn:specieseqjump} reduces to \cite{irfan2017front}
\begin{equation} 
\begin{aligned}
{\dot{m}}_{\Gamma}=\frac{{\rho }_g\ D_{vg}\ {\left(\frac{\partial Y}{\partial n}\right)}_{\Gamma}}{1-Y^{\Gamma}},
\end{aligned}
\label{eqn:speciessingle}
\end{equation}
where ${\left(\frac{\partial Y}{\partial n}\right)}_{\Gamma}$ is the mass fraction gradient evaluated at the interface in the normal direction. The vapor saturation mass fraction at the interface, $Y^{\mathit{\Gamma}}$, is related to the vapor pressure, $P^{\Gamma}_v$, as
\begin{equation} 
\begin{aligned}
Y^{\Gamma}=\frac{P^{\Gamma}_vM_v}{P^{\Gamma}_vM_v+{(P_{\rm atm}-P}^{\Gamma}_v)\ M_g},
\end{aligned}
\label{eqn:Y-Clasius}
\end{equation}
where $P_{\rm atm}$ is the ambient pressure taken here as the atmospheric pressure. The vapor pressure can be obtained from the Clasius-Clapeyron relation as
\begin{equation} 
\begin{aligned}
P^{\Gamma}_v=P_{atm}\ {\mathrm{exp}\left(-h_{lg}\frac{M_v}{\cal R}\ \left(\frac{1}{T_{\Gamma}}-\ \frac{1}{T_b}\right)\right)},
\end{aligned}
\label{eqn:P-Clasius}
\end{equation}
where $P^{\Gamma}_v$ is the saturated vapor pressure at the interface temperature $T_{\Gamma}$, $T_b$ is the boiling temperature of the liquid at $P_{atm}$, $M_v$ and $M_g$ are, respectively, the molar masses of the vapor and gas, and ${\cal R}$ is the universal gas constant. 

It is further assumed that the material properties remain constant following a fluid particle:
\begin{equation} 
\begin{aligned}
\frac{D\rho }{Dt}=0;\ \frac{D\mu }{Dt}=0;\ \frac{Dk}{Dt}=0;\ \frac{Dc_p}{Dt}=0;\ \frac{DD_{vg}}{Dt}=0,
\end{aligned}
\label{eqn:constprop}
\end{equation}
where $\frac{D }{Dt}=\frac{\partial}{\partial t} + \boldsymbol{u}\cdot\nabla$ is the material derivative. The material properties are set in the entire computational domain using an indicator function, $I$. Letting $\Psi$ denote the set of all material properties, it is computed as
\begin{align}
\begin{split}
&\Psi ={\Psi }_l\ I\left(\boldsymbol{x},t\right)+\ {\Psi }_g\ \left(1-I\left(\boldsymbol{x},t\right)\right),  
\end{split}
\label{eqn:thermophysical}
\end{align}
where the indicator function is defined as:
\begin{equation} 
I\left(\boldsymbol{x},t\right)=\ \left\{ \begin{array}{ll}
1 &\ {\rm in}\ {\rm the}\ {\rm droplet}\ {\rm phase}, \\ 
0& \ {\rm in}\ {\rm the}\ {\rm bulk}\ \ {\rm phase}. 
\end{array}
\right.
\end{equation}
Note that the vapor mass fraction evolution equation, (i.e., Eq.~(\ref{eqn:specieseq})) is solved only in the bulk phase, so the binary diffusion coefficient, $D_{vg}$, is defined only outside of the droplet and the associated interfacial boundary condition is enforced using the immersed-boundary method as will be discussed in Section~\ref{IPGC}.
\section{Numerical Method}
\label{NumMeth}
\begin{figure}[!t]
\centering
\includegraphics[scale=1.25]{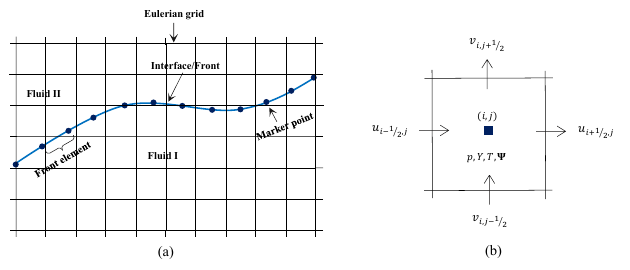}
\caption{(a) A schematic representation of the Lagrangian grid cast on the stationary Eulerian grid. (b) The staggered grid arrangement used to solve the field equations.} 
\label{Fig:FTillust}
\end{figure}

In the finite-difference/front-tracking method \citep{unverdi1992front,tryggvason2001front,juric1998computations,esmaeeli2004computations,irfan2018front}, a stationary staggered Eulerian grid is used to solve the flow equations fully coupled with the energy and vapor mass-fraction evolution equations while the interface is represented by a separate Lagrangian grid as shown in Fig.~\ref{Fig:FTillust}. In this study, the flow and energy conservation equations are solved in the entire computational domain using a one-field formulation but the species mass-fraction equation is solved only in the ambient fluid and a sharp-interface immersed-boundary method is used to impose the boundary conditions at the interface. The Lagrangian grid (also called front) consists of connected marker points that move with the combination of the local flow velocity interpolated from the neighboring  Eulerian grid and the velocity induced by the phase change (see Eq. \eqref{eqn:normalv}). A Lagrangian element between two adjacent marker points is called a front element. The Lagrangian grid cast on the Eulerian grid is depicted schematically in Fig.~\ref{Fig:FTillust}a. In the staggered grid arrangement, the velocity components are placed on the cell faces while the scalar quantities such as pressure, temperature, mass fraction and all the material properties are stored at the cell centers as shown in Fig.~\ref{Fig:FTillust}b. 

The Lagrangian marker points are advected in the normal direction as 
\begin{equation} 
\begin{aligned}
\frac{d{\boldsymbol{x}}_{\Gamma}}{dt}=u_n{\boldsymbol{n}}_{\Gamma},
\end{aligned}
\label{eqn:frontadv}
\end{equation}
where $\boldsymbol{x}_{\Gamma }$ is position of marker points, $\boldsymbol{n}_{\Gamma }$ is the outward normal vector and  $u_n$ is the normal component of the velocity computed as
\begin{equation} 
\begin{aligned}
u_n=\frac{1}{2}\left({\boldsymbol{u}}_l+{\boldsymbol{u}}_g\right) \cdot \boldsymbol{n}-\frac{\dot{q_{\Gamma}}}{2 h_{lg }}\left(\frac{1}{{\rho }_l}+\frac{1}{{\rho }_g}\right),
\end{aligned}
\label{eqn:normalv}   
\end{equation}
where $\boldsymbol{u}_l$ and $\boldsymbol{u}_g$ are the liquid and gas phase velocities, respectively, and they are evaluated at the interface using one sided interpolation. Full details of the front-tracking method can be found in the review paper by \citet{tryggvason2001front} and the book by \citet{tryggvason2011direct}. 
\subsection{Flow solver} 
The flow equations are solved on the staggered Eulerian grid using a projection method \citep{chorin1968numerical} in which the momentum equation is first written in a semi-discretized form as
\begin{equation} 
\rho^{n}\frac{{\boldsymbol{u}}^{n+1}- {\boldsymbol{u}}^n}{\Delta t}={\boldsymbol{A}}^n- \nabla p,
\label{eqn:projection0}
\end{equation}
where $\Delta t$ is the time step and $\boldsymbol{A}$ represents the convective, diffusive and body force terms. The superscript $n$ denotes the current time step. This equation is solved in two steps:
\begin{eqnarray} 
\rho^{n} \frac{{\boldsymbol{u}}^*-{\boldsymbol{u}}^n}{\Delta t}&=&{\boldsymbol{A}}^n,  \label{eqn:projection1} \\
\rho^{n} \frac{{\boldsymbol{u}}^{n+1}-{\boldsymbol{u}}^*}{\Delta t} &=&- \nabla p, \label{eqn:projection2}
\end{eqnarray}
where ${\boldsymbol{u}}^*$ is the unprojected velocity. Taking a divergence of Eq.~(\ref{eqn:projection2}) results in a non-separable Poisson equation for the pressure in the form
\begin{eqnarray}
\nabla \cdot \frac{1}{{\rho }^{n\ }} \nabla p= \frac{\nabla \cdot {\boldsymbol{u}}^*- {\nabla \cdot \boldsymbol{u}}^{n+1}}{\Delta t},
\label{eqn:Poisson}
\end{eqnarray}
where ${\nabla \cdot \boldsymbol{u}}^{n+1}$ is computed employing Eq.\eqref{eqn:continuity} as
\begin{equation} 
\nabla \cdot \boldsymbol{u}^{n+1}=\left[\frac{1}{h_{lg}}\left(\frac{1}{{\rho }_g}-\frac{{1}}{{\rho }_l}\right)\int_A{\delta \left(\boldsymbol{x}-{\boldsymbol{x}}_{\mathit{\Gamma}}\right)\dot{q}_{\Gamma} \ dA}\right]^{n+1}.
\label{eqn:continuity2}
\end{equation}
The pressure Poisson equation (Eq.~(\ref{eqn:Poisson})) is solved using a multigrid method implemented in MUDPACK \cite{MUDPACK} package as discussed by \citet{tryggvason2001front}. Once the pressure field is computed, the velocity field is then corrected to satisfy the mass conservation as
\begin{equation}
\boldsymbol{u}^{n+1} = \boldsymbol{u}^{*} - \frac{\Delta t}{\rho^n}\nabla p.
\end{equation}
In Eqs.~(\ref{eqn:projection0})-(\ref{eqn:Poisson}), the spatial derivatives in the convective terms are evaluated using a third-order QUICK scheme \cite{LEONARD197959} while all other spatial derivatives are discretized using central differences on the staggered grid. The surface tension is computed on the interface at the center of the front elements and it is then distributed onto the neighboring Eulerian grid points conservatively in the same manner as described by \citet{tryggvason2001front,tryggvason2011direct}.

The energy and mass-fraction equations are discretized as
\begin{eqnarray} 
\frac{T^{n+1}-T^{n}}{\Delta t} &=& \left[ \frac{1}{\rho c_p} \left(-\boldsymbol{u}\cdot\nabla T + \nabla \cdot k \nabla T- \int_A{\delta \left(\boldsymbol{x}-{\boldsymbol{x}}_{\Gamma}\right)\dot{q}_{\Gamma} \ dA}\right) \right]^{n},
\label{eqn:energyeq2} \\
\frac{Y^{n+1}-Y^{n}}{\Delta t}&=& \left[-\boldsymbol{u} \cdot \nabla Y + \nabla \cdot D_{vg}\nabla Y\right]^{n},  
\label{eqn:specieseq2}
\end{eqnarray}
where the convective terms are approximated using a fifth-order WENO-Z scheme \cite{borges2008improved} while all other spatial derivatives are evaluated using the central differences on the staggered grid.

The numerical method described above is explicit and only first order in time. However, a formally second-order accuracy can be easily achieved using a predictor-corrector scheme as described by \citet{tryggvason2001front}. Although a second-order predictor-corrector scheme is implemented, following \citet{irfan2018front}, the first order method is used in the present simulations since the temporal discretization error is generally found to be negligibly small compared to the spatial error owing to a small time step imposed by the numerical stability.

\subsection{Interfacial Boundary Conditions: Sharp-Interface Immersed-Boundary Method}
\label{IPGC}
\begin{figure}[ht]
\centering
\includegraphics[scale=0.95]{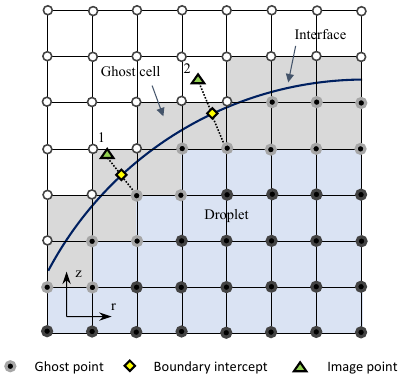}
\caption{Schematic representation of the sharp-interface immersed-boundary methodology used to impose the species mass boundary condition at the interface. GP, BI and IP denote a ghost cell, a boundary-intercept point, and an image-point, respectively. The the image points denoted by 1 and 2 are located partially and totally  in the bulk fluid cell, respectively.}
\label{Fig:IBMsketch}
\end{figure}

The mass-fraction is solved only in the ambient fluid so a special treatment is required to impose the boundary conditions at the interface for $Y$. In the present work, we use the image point and ghost cell methodology developed by \citet{mittal2008versatile} to accurately impose the Dirichlet boundary condition for the mass fraction at the droplet surface. Note that the same methodology is also used to simulate the flow over a solid sphere in a convective environment as will be discussed in Section \ref{IBM}. Following \citet{ZolfaghariEtAl-17-caf}, Eulerian cells cut through by the Lagrangian grid are identified and called ghost cells (GC) as sketched in Fig.~\ref{Fig:IBMsketch}. Similarly, the vertices of the ghost cells inside the droplet region are called ghost points (GP). Then the boundary intercept (BI) points are determined by drawing a normal line from the associated ghost points to the Lagrangian grid. Finally, the image point (IP) is computed by extending the normal line into ambient fluid by the same distance. 
After identifying all the BIs and the associated IPs, the vapor mass fraction, $Y$, is represented in the cell containing an IP by a bilinear interpolant in the form

\begin{equation} 
\begin{aligned}
Y \left(r,z\right)=a_1rz+a_2r+a_3z+a_4.
\end{aligned}
\label{eqn:ipgc-interpolation}
\end{equation}
If the cell containing IP is totally in the ambient fluid domain, the mass fraction at the image point, $Y_{IP}$, is simply computed by a bilinear interpolation. Otherwise, a linear system is formed to compute the interpolation coefficients as
\begin{equation} 
\begin{aligned}
\left[V\right]\left\{A\right\}=\left\{Y \right\},
\end{aligned}
\label{eqn:ipgc-phi}
\end{equation}
where 
\begin{equation} 
\begin{aligned}
\left[V\right]=\left[ \begin{array}{cccc}
r_1z_1 & r_1 & z_1 & 1 \\ 
r_2z_2 & r_2 & z_2 & 1 \\ 
r_3z_3 & r_3 & z_3 & 1 \\ 
r_4z_4 & r_4 & z_4 & 1 \end{array}
\right]; \;
\left\{A\right\}=\left\{ \begin{array}{c} a_1\\ a_2\\ a_3\\ a_4\end{array} \right\}; \;
\left\{Y\right\}=\left\{ \begin{array}{c} Y_1\\ Y_2\\ Y_3\\ Y_4\end{array} \right\}. 
\end{aligned}
\label{eqn:ipgc-Vandermonde}
\end{equation}
In Eq.~(\ref{eqn:ipgc-Vandermonde}), the entries corresponding to a GP are replaced with the boundary conditions. For instance, suppose that the second node is a GP, then we set $r_2 = r_{BI}, z_2 = z_{BI}$ and $Y_2=Y_{BI}$. Note that $Y_{BI}$ is computed at the interface using Eq.~(\ref{eqn:Y-Clasius}). Once the interpolation coefficients are determined from Eq.~(\ref{eqn:ipgc-phi}), then mass fraction at the image point is simply computed as $Y \left(r_{IP},z_{IP}\right) = a_1 r_{IP} z_{IP}+a_2 r_{IP}+a_3 z_{IP}+a_4$. Finally, the ghost cell value is computed as
\begin{equation} 
\begin{aligned}
Y_{GC}=2Y_{BI}-Y_{IP}.
\end{aligned}
\label{eqn:Dirichlet}
\end{equation}
Note that as a fifth-order WENO-Z scheme \cite{borges2008improved} scheme is used to estimate the convective fluxes in Eq.~\eqref{eqn:specieseq}, we employ a larger bandwidth around the interface to locate the ghost and image points as depicted in Fig.~\ref{Fig:ibmillust}b. The details of the present sharp-interface immersed-boundary method can be found in \citet{mittal2008versatile} and \citet{ZolfaghariEtAl-17-caf}.

\subsection{Species-based evaporation model}
\begin{figure}[ht]
\centering
\includegraphics[scale=0.45]{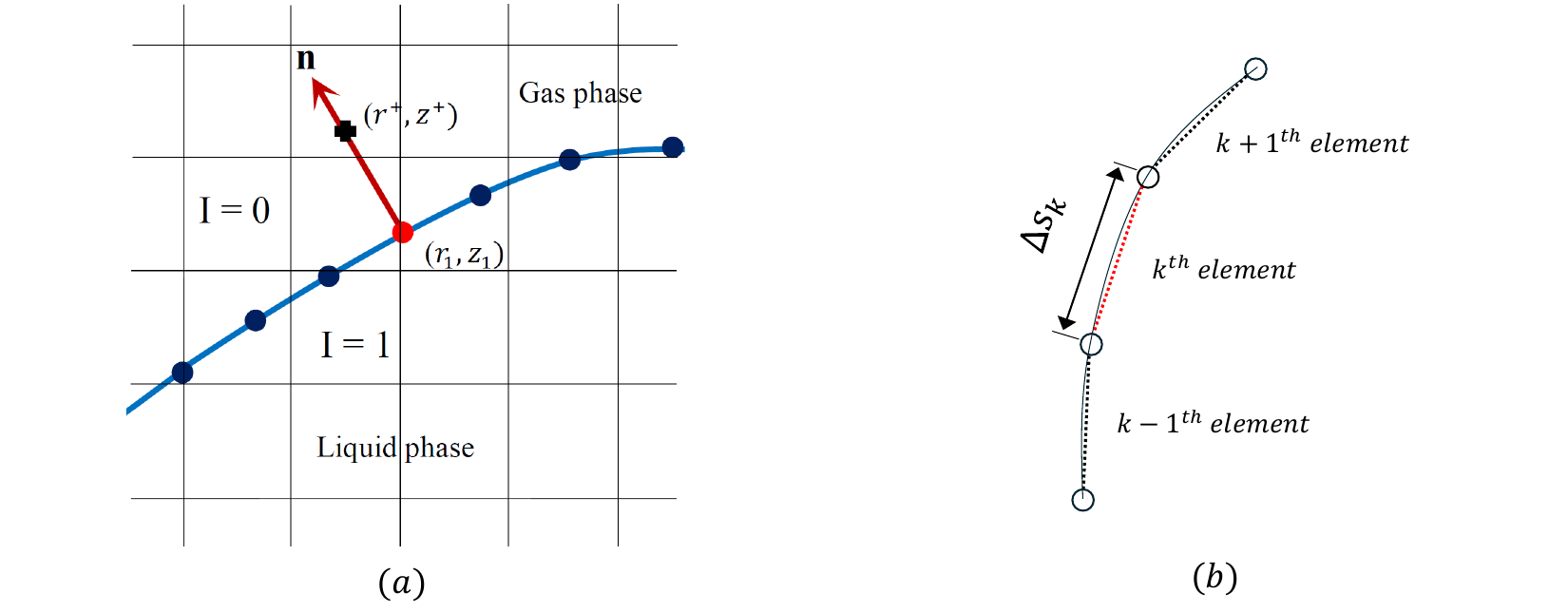}
\caption{(a) The procedure to find the mass fraction gradient at the interface. (b) A schematic illustration of a front element and its length, $\Delta s_k$. }
\label{Fig:yplus}
\end{figure}

The mass flux is computed at the center of the front element from Eq.~\eqref{eqn:speciessingle} as 
\begin{equation} 
\begin{aligned}
{\dot{m}}_{\Gamma_k}=\left[\frac{{\rho }_g\ D_{vg}\ {\left(\frac{\partial Y}{\partial n}\right)}_{\Gamma}}{1-Y^{\Gamma}}\right]_k.
\end{aligned}
\label{eqn:speciessingle2}
\end{equation}
Following \citep{irfan2017front,irfan2018front}, the mass fraction gradient in the gas phase normal to the interface is approximated as
\begin{equation} 
\begin{aligned}
{\left(\frac{\partial Y}{\partial n}\right)}_{\Gamma}=\frac{1}{\eta h }\ \left(Y^+-Y^{\Gamma}\right).
\end{aligned}
\label{eqn:sepcnormal}
\end{equation}
Referring to Fig.~\ref{Fig:yplus}, the point $(r^+, z^+)$ is found in the gas phase at a distance $\eta h$ from the front element centroid where $h$ is the Eulerian grid spacing and $\eta$ is a parameter set to 2 in the present study. The normal vectors are easily obtained using  smooth tangent vectors that are already calculated using third order Legendre polynomials at each marker point to find the curvature \cite{tryggvason2011direct} in computing the surface tension. Once $(r^+, z^+)$ is determined, the mass fraction, $Y^+$, is approximated using a bi-linear interpolation. Using the mass flux value, the heat flux is computed as
\begin{equation} 
\begin{aligned}
\dot{q}_{\Gamma}=h_{lg}{\dot{m}}_{\Gamma},
\end{aligned}
\label{eqn:energynormal}
\end{equation}
which is conservatively distributed onto neighboring Eulerian grid cells to be added as a source term to the energy and continuity equations. The conservation requires
\begin{equation} 
\begin{aligned}
\int_{\Delta A}{{\dot{q}_{\Gamma}}\left(A\right)dA=} \int_{\Delta V}{\dot{q}\left(\boldsymbol{r}\right) dV},
\end{aligned}
\label{eqn:smooth}
\end{equation}
where $\dot{q}$ is the grid value. Following Tryggvason et al.~\citep{tryggvason2001front,tryggvason2011direct}, $\dot{q}_{\Gamma}$ is smoothed on the Eulerian grid as
\begin{equation} 
\begin{aligned}
\dot{q}_{i,j} = \sum_k \dot{q}_{\Gamma }^k w^k_{i,j} \frac{r_k\Delta s_k}{r_{i,j}h^2},
\end{aligned}
\label{eqn:phiformula}
\end{equation}
where $\mathrm{\Delta }s_k$ is the length of the front element, $r_k$ is the radial coordinate of the element center, $r_{i,j}$ is the radial position of the respective grid node and $w^k_{i,j}$ is the weight of the Eulerian grid node $(i,j)$ corresponding to the $k^{th}$ element. The weight is calculated using the Peskin’s cosine function \citep{peskin1977numerical,tryggvason2001front,irfan2018front}.

\subsection{Restructuring the Lagrangian grid}
\begin{figure}[ht]
\centering
\includegraphics[scale=0.35]{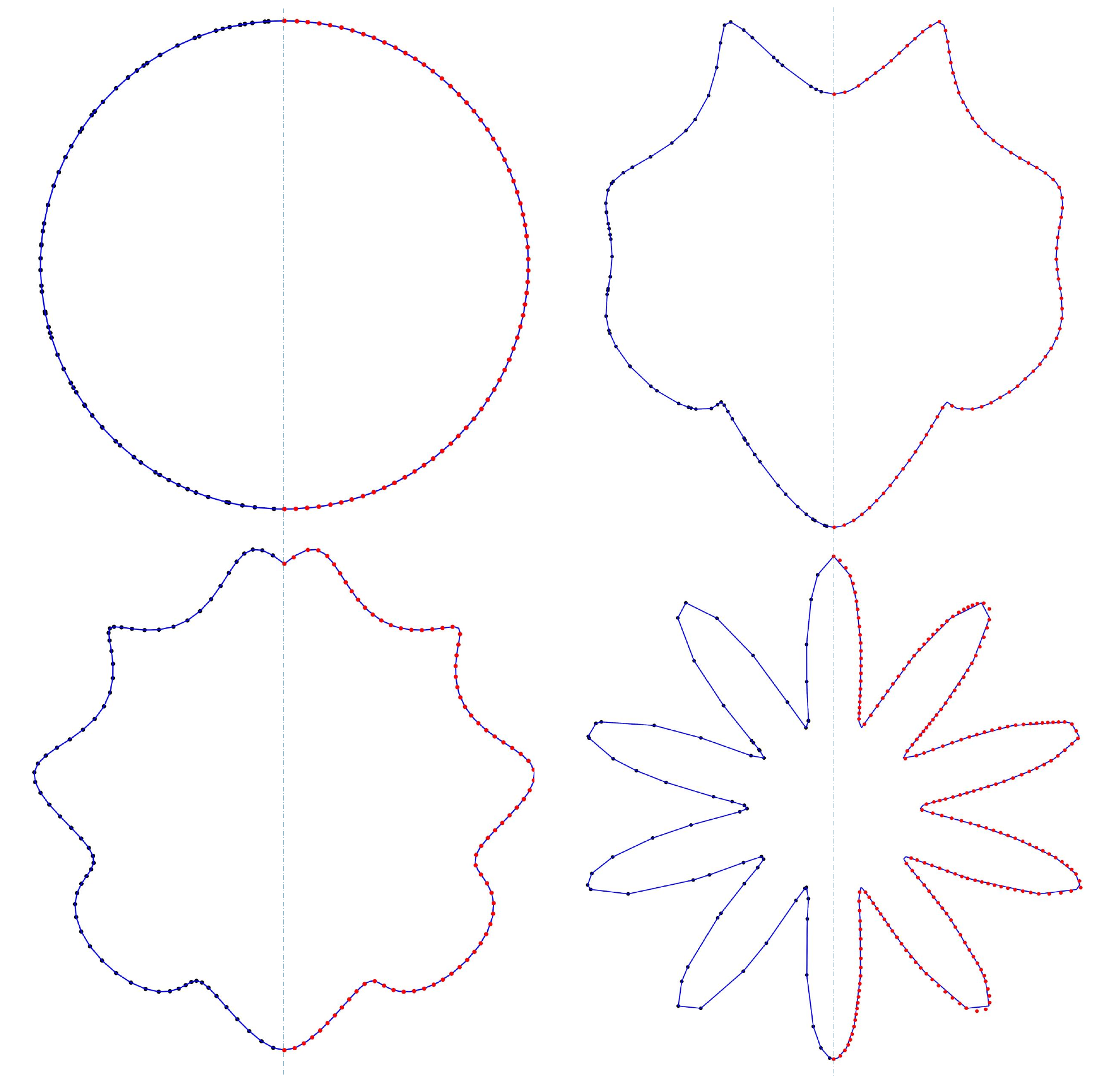}
\caption{Performance of the new restructuring procedure for the Lagrangian grid. The randomly initialized (black dots on left side) and uniformly restructured (red dots on right side) front marker points for various shapes. The solid blue lines show the prespecified interfaces.}
\label{Fig:4shapes}
\end{figure}

As the interface continuously deforms and stretches, the Lagrangian grid may become highly non-uniform resulting in an inadequate resolution in some parts while overcrowded in other parts (\citet{tryggvason2001front}). It is thus needed to restructure the Lagrangian grid dynamically to keep the front element size nearly uniform throughout the simulation. In the present study, we use a slightly modified version of the original restructuring algorithm developed by \citet{tryggvason2001front}. The new procedure is designed to keep  the Lagrangian grid uniform throughout a simulation.  The algorithm utilizes the same third order Legendre polynomial fit used to compute the interfacial curvature \cite{tryggvason2001front}. Starting with the first marker point located on the centerline, we draw a circle of a prespecified radius centered at  the marker point and locate the next marker point at the intersection of the circle and the  Legendre polynomial, and repeat this procedure until all the marker points are located. This simple algorithm preserves the interface curvature and locates front marker points evenly distributed around droplet perimeter. However, a special treatment may be needed for the second-to-last marker point to avoid the last front element being too small near the centerline. In such cases, the marker point is either removed or repositioned to ensure that the size of the last front element remains within the acceptable threshold. Figure~\ref{Fig:4shapes} demonstrates the performance of the new front restructuring algorithm for various shapes. The marker points are initialized randomly on the shapes. As seen, the new restructuring algorithm adeptly redistributes the marker points uniformly while preserving the shape of the droplet and its curvature.

\subsection{Overall solution algorithm}
\label{algorithm}
The overall solution algorithm can be summarized as follows:
\begin{enumerate}
    \item The ghost cells, the boundary intercepts, and the image points are determined using a set of computational geometry operations.
    \item 	The heat flux is computed using Eq.~\eqref{eqn:energynormal} at the $n^{th}$ time step  to account for any newly created marker points, and is distributed conservatively on the fixed Eulerian grid. 
    \item  The saturation mass fraction boundary condition is set as a Dirichlet boundary condition on the interface using local temperature interpolated from neighboring Eulerian grid nodes at time level $n$.
    \item 	The Lagrangian marker points are advected according to Eq.\eqref{eqn:frontadv} to determine the location of the interface at the new time step, i.e., ${x_{\Gamma}}^{n+1}={x_{\Gamma}}^n + \Delta t {(u_n  n_{\Gamma})}^n$.
    \item The indicator function is computed based on the new location of the marker points and all the material properties are updated according to Eq.~\eqref{eqn:thermophysical}.
    \item The energy and species equations are solved to obtain $T^{n+1}$ and $Y^{n+1}$.
    \item The heat source term at n+1 time level is evaluated using the updated energy field and front location, and it is distributed on the fixed grid to be used as a source term in the pressure Poisson equation.
    \item The pressure Poisson equation, Eq.~\eqref{eqn:Poisson}, is solved, and the velocities at the new time step are corrected. 
    \item The Lagrangian grid is restructured if needed. 
\end{enumerate}

\section{Results and Discussion}
\label{results}
\subsection{Validation of Multiphase Flow Solver}
\label{MultiphaseValid}
\begin{figure}[!t]
\centering
\includegraphics[scale=0.72]{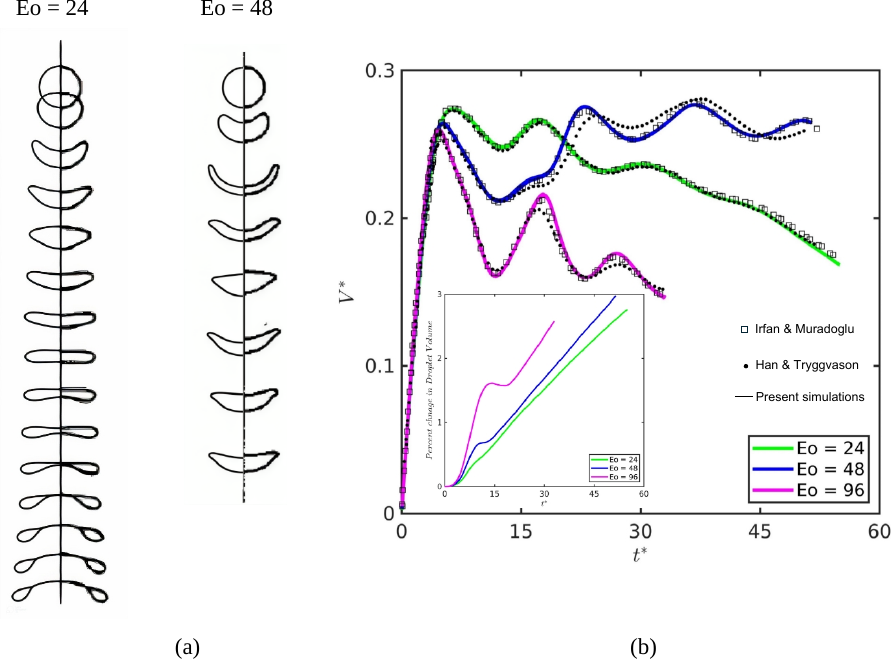}
\caption{(a) The evolution of the droplet shape for $Eo = 24$ and $Eo = 48$ cases. The present results (left side) are compared with the results of \citet{han1999secondary} (right side). The time interval between two successive drops in each column is $\Delta t^* = 3.953$ and $\Delta t^* = 5.59$ for $Eo = 24$ and $Eo = 48$, respectively. (b) Comparison of the non-dimensional centroid velocity of a falling droplet with the results of \citet{irfan2018front} and \citet{han1999secondary} for $Eo= 24, 48$ and 96. The inset shows the percentage change of the droplet volume during the simulation. ($Oh_o=0.05, Oh_i=0.0466, \rho_i/\rho_o =1.15, \mu_i/\mu_o=1$, Grid: $512\times1536$) }
\label{Fig:trygcomp}
\end{figure}

We first perform simulations for a standard test case of a non-evaporating gravity-driven falling droplet \cite{han1999secondary,irfan2018front} to validate the multiphase solver as well as the new Lagrangian grid restructuring algorithm. The computational domain is $5d\times 15d$ and it is resolved by a uniform Cartesian grid of $512\times 1536$ grid cells in the radial and axial directions, respectively. A spherical droplet of diameter $d$ is initialized at $(r_c,z_c)=(0,13.75d)$ in a rigid cylinder filled with an otherwise quiescent ambient fluid. The no-slip and axisymmetry boundary conditions are applied at the cylinder wall and at the centerline, respectively. Simulations are performed for the density ratio of $\rho_i/\rho_o = 1.15$, the viscosity ratio of $\mu_i/\mu_o = 1$ and the Ohnesorge numbers of $Oh_o = \mu_d/\rho_o d\sigma = 0.05$ and $Oh_i = \mu_i/\rho_i d\sigma = 0.0466$, where subscripts $'o'$ and $'i'$ denote the continuous and dispersed phases, respectively. The E\"{o}tv\"{o}s number is defined as $Eo = g_z\left(\rho_i-\rho_o\right) d^2/\sigma$ and the simulations are repeated for the $Eo=24$, $Eo=48$ and $Eo=96$ cases. The results are shown in Fig.~\ref{Fig:trygcomp} in the non-dimensional form using the length scale $d$, time scale $\sqrt{d/g_z}$ and velocity scale $\sqrt{dg_z}$. As seen, the present results are in excellent qualitative and quantitative agreement with the published results of \citet{irfan2018front} demonstrating the accuracy of the present multiphase solver. Note that the slight discrepancy with the results of \citet{han1999secondary} is attributed to the lack of resolution in their simulation \cite{han1999secondary}. Since flow is incompressible, droplet volume change occurs solely due to accumulation of numerical error. The percentage change in the droplet volume is plotted in the inset of Fig.~\ref{Fig:trygcomp}b. The volume conservation error is found to be less than 3\% for all the cases.

\subsection{Mass Transfer from a Solid Sphere}
\label{IBM}
\begin{figure}[ht]
\centering
\includegraphics[scale=0.40]{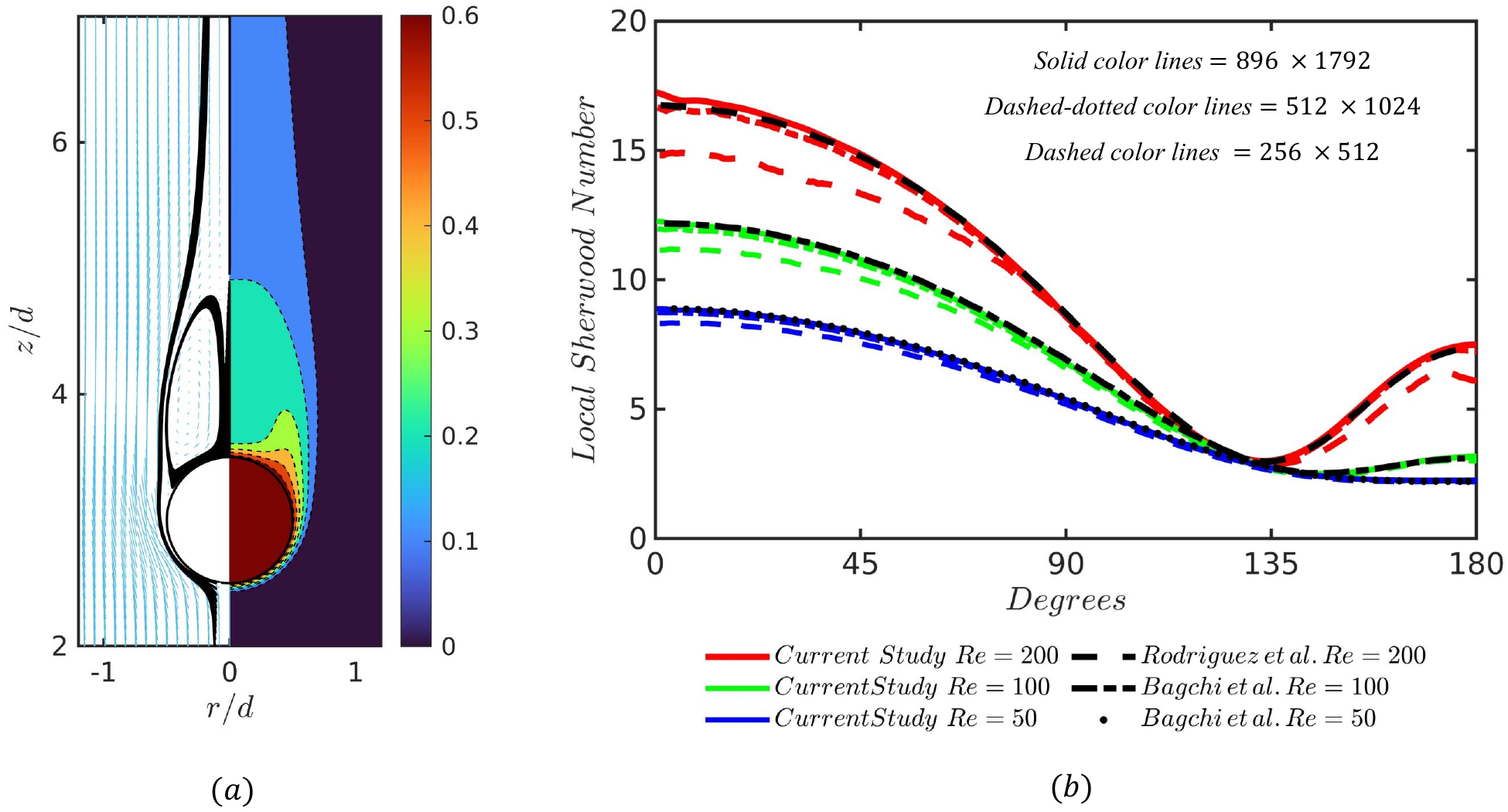}
\caption{(a) Velocity vectors and streamlines (left portion) and constant contours of species mass fraction (right portion) around a solid sphere at $Re=200$. (b) Variation of the local Sherwood number. The present results are compared with those of \cite{bagchi2001direct} for  $Re = 50$ and $Re=100$, and of \cite{rodriguez2018boundary} for $Re=200$. Colored dashed, dashed-dotted, and solid lines indicate the present results computed on the grid resolutions of $256\times 512$, $512\times 1024$ and $896\times 1792$, respectively. ($Sc=0.7$.)}
\label{Fig:solidsphere}
\end{figure}

The sharp-interface immersed-boundary (IBM) method is first tested for the case of mass transfer from a solid sphere of diameter $d$ immersed in a uniform ambient flow at moderate Reynolds numbers in the range $50\le Re\le 200$.  For this purpose, the IBM method of \citet{mittal2008versatile} is used to impose the no-slip velocity boundary conditions on the solid sphere surface in addition to the Dirichlet boundary condition applied for the mass transfer. The computational domain is $4d\times 8d$ in the radial and axial directions, respectively. The sphere is held stationary with its center being located at $3d$ from the inlet in the axial direction.  A uniform velocity $U_{\infty}$ is applied at the inlet while constant pressure and zero axial velocity gradient boundary conditions are applied at the outlet. The gradient-free boundary conditions are imposed at the far-field, i.e., at $r=4d$.  The simulations are performed using a $256\times 512$ (coarse), $512\times1024$ (moderate), and $896\times1792$ (fine) grid resolutions to demonstrate the grid convergence. Figure~\ref{Fig:solidsphere}a illustrates the velocity vectors, streamlines, and the mass fraction field around the sphere at $Re=200$. As seen, a bound vortex is created in the wake region behind the sphere. The  length of the recirculation zone is computed as $L_w = 0.45d$, $0.9d$, $1.43d$ for $Re=50$,$Re=100$, and $Re=200$, respectively. These values are in good agreement with the experimental values of $0.48, 0.85$ and $1.2$ \citep{KalraUhlherr_ces_73,clift2005bubbles} and the numerical results of $0.4, 0.87$ and $1.42$  \citep{TOMBOULIDES_ORSZAG_2000}.  Figure~\ref{Fig:solidsphere}b shows the local Sherwood number a function of the angle measured from the stagnation point at the leading edge. The Sherwood number is defined as:
\begin{equation}
\begin{aligned}
    Sh = \frac{-2R}{Y^\Gamma - Y_\infty} \left(\frac{\partial Y}{\partial n}\right)_\Gamma,
\end{aligned}
\label{eqn:u}
\end{equation}
where $R$ is the sphere radius and the gradient term is evaluated in the same way as in Eq.~\eqref{eqn:sepcnormal}. Simulations are repeated for three different grid resolutions and the results are compared with the computational results of \citet{bagchi2001direct} for $Re=50$ and $Re=100$, and of \citet{rodriguez2018boundary} for $Re=200$. As seen, the local Sherwood number is captured almost perfectly using the moderate and fine grid resolutions except for the front stagnation point at higher Reynolds numbers. It is worth noting that the mild wiggles observed near the stagnation point are attributed to the rapid circumferential variation of the distance between the boundary intercept points where the boundary conditions are imposed and the corresponding ghost points, as also previously reported by \citet{majumdar2001rans}. We note that the computations are performed for the Reynolds number up to $Re=200$ since a vortex shedding is expected to occur behind the sphere for $Re\gtrsim 210$ \citep{bagchi2001direct}, which cannot captured by the present axisymmetric simulations. 

\subsection{The \texorpdfstring{$d^2$}{d-squared}-law}
\label{d2law}
\begin{figure}[!t]
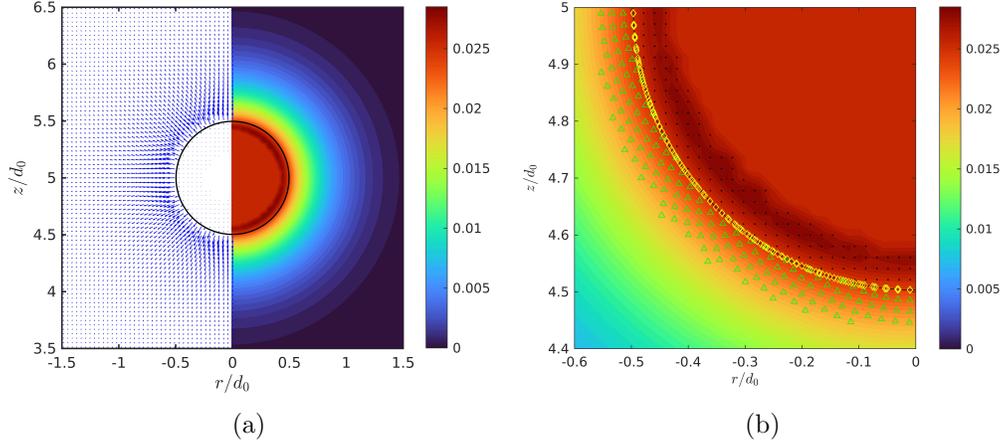

\centering
\begin{tabular}{cc}
\includegraphics[scale=0.024]{./d2BM0025best.pdf} 
&
\includegraphics[scale=0.024]{./IBMillust.pdf} \\
\footnotesize (a) & \footnotesize (b)
\end{tabular}
\caption{(a) Velocity vectors (left side) and mass fraction field (right side) around an evaporating droplet. (b) An enlarged view illustrating the vapor mass fraction contours as well as distribution of the ghost points ($GP$), boundary intercepts ($BI$) and image points ($IP$) in vicinity of the interface indicated by black dots, yellow diamonds and green triangles, respectively. Constant contours represent the mass fraction field. $\left(B_M = 0.025,\; t^*=0.7\right)$. }
\label{Fig:ibmillust}
\end{figure}

We next perform simulations for the well-known test case of a stationary single droplet evaporating in an stagnant ambient air. The initial diameter of the droplet is $d_0$. The buoyancy-induced natural convection is neglected. The temporal evolution of the droplet diameter is given by \cite{turns2000introduction,williams2018combustion}:
\begin{equation}
\begin{aligned}
    \frac{dd^2}{dt} = -8 \frac{\rho_g D_{vg}}{\rho_l} ln(1+B_M),
\end{aligned}
\label{eqn:dsquare}
\end{equation}
where $\rho_g$, $\rho_l$ and $D_{vg}$ denote the gas density, the liquid density and the diffusion coefficient of the vapor in the surrounding air, respectively. The mass transfer potential is defined as $B_M=(Y^{\Gamma}-Y_{\infty })/({1-Y^{\Gamma}})$ where $Y^{\Gamma}$ and $Y_{\infty}$ denote the saturation and the far-field vapor mass fractions, respectively. The saturation mass fraction, $Y^{\Gamma}$, is set as a Dirichlet boundary condition on the droplet surface. Simulations are performed for a droplet centered at $(0,5d_0)$ in a computational domain of $5d_0\times 10d_0$ using a $256\times 512$ uniform grid resolution in the radial and axial directions, respectively. The time and length are scaled by $d_0$ and $\frac{d_0^2}{D_{vg}}$, respectively. Figure~\ref{Fig:ibmillust}a illustrates the velocity vectors (Stefan flow) and the constant contours of the vapor mass fraction in the vicinity of the droplet for the case of $B_M = 0.025$ at the scaled time of $t^* = 0.7$. The smooth flow and mass fraction fields qualitatively indicate the accuracy of the numerical method. The enlarged view in Fig.~\ref{Fig:ibmillust}b shows the vapor mass fraction as well as the distribution of the ghost cells, the boundary intercepts, and the image points near the interface. 


\begin{figure}[!t]
\centering
\includegraphics[scale=0.435]{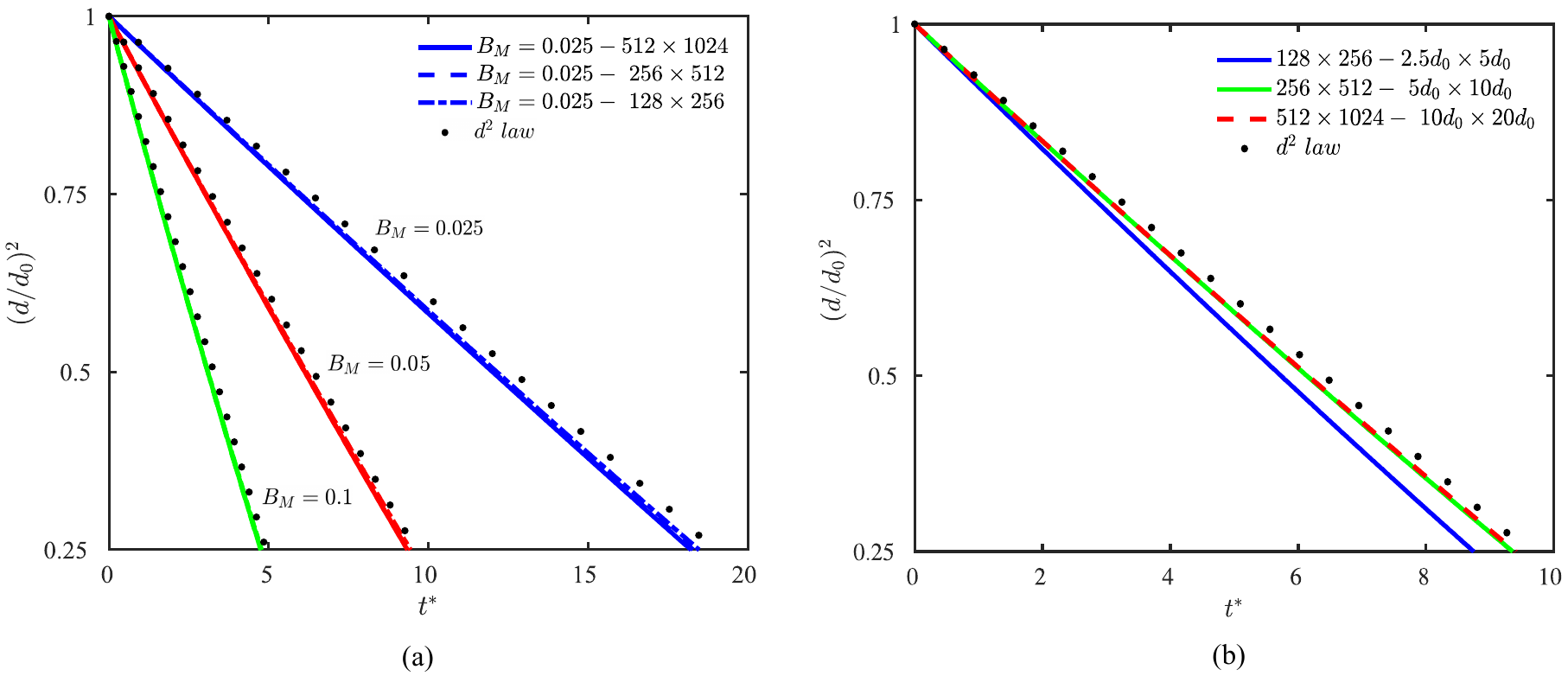}
\caption{(a) Validation of the numerical method for the $d^2$ law. The numerical results are compared with the analytical solutions for the mass transfer numbers of $B_M = 0.025,\; B_M = 0.05$, and $B_M = 0.1$. Simulations are performed using $128\times 256$, $256\times 512$ and $512\times 1024$ grid resolutions to demonstrate the grid convergence. Domain size: $5d_0 \times 10d_0$. (b) The effects of the domain size on the numerical results and comparison with the $d^2$-law. The simulations are performed for the domain sizes of $2.5d_0\times 5d_0$, $5d_0\times 10d_0$ and $10d_0\times 20d_0$ using $128\times 256$, $256\times 512$, $512\times 1024$ grid resolutions, respectively. ($B_M = 0.05$).}
\label{Fig:d2species}
\end{figure}

\begin{figure}[ht]
\centering
\includegraphics[scale=0.48]
{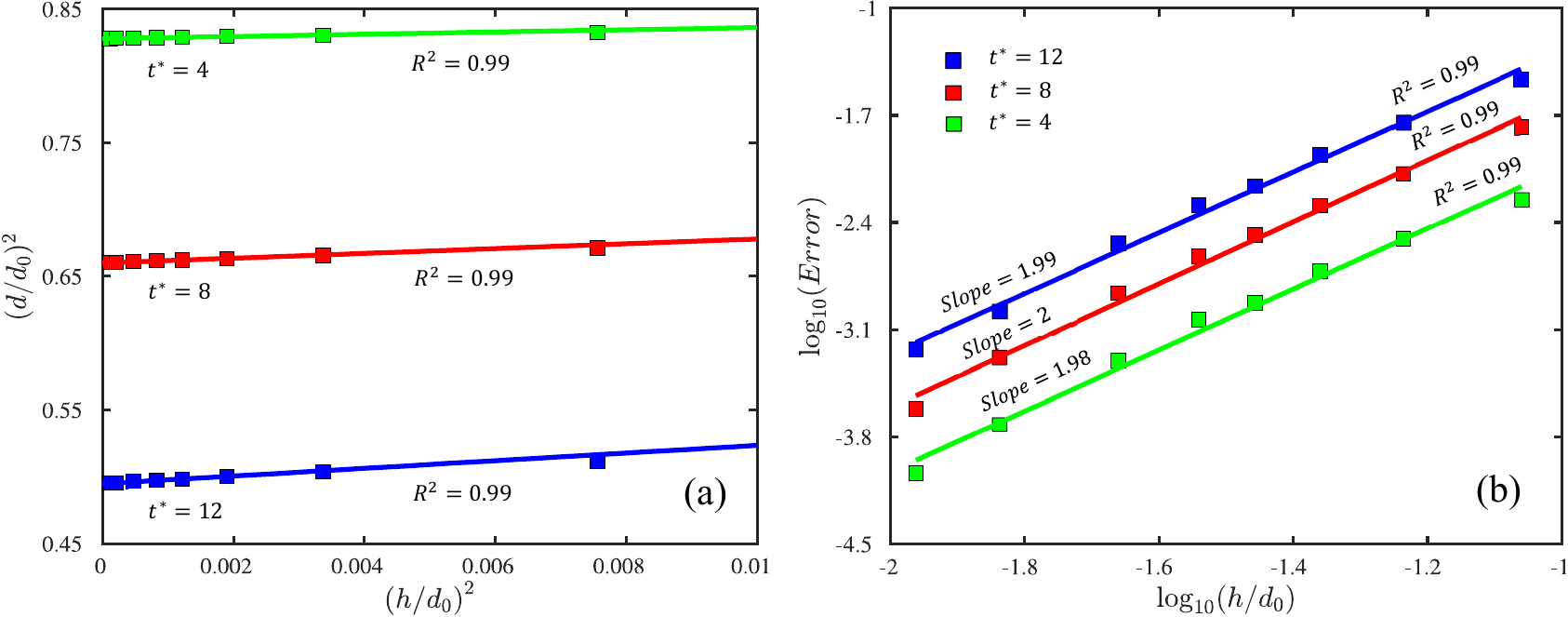}
\caption{Quantification of the spatial accuracy. The variation of (a) $(d/d_0)^2$  and (b) the spatial error against $(h/d_0)^2$ for a droplet evaporating in stagnant condition. Simulations are performed on the domain size of $5d \times 10d$ for $B_M = 0.025$ using $64\times 128$, $96\times 192$, $128\times 256$, $160\times 320$, $192\times 384$, $256\times 512$, $384\times 768$ and $512\times 1024$ and the results are taken at $t^* =4,\; t^* = 8$ and $t^* = 12$. The symbols and the solid lines indicate the numerical results and the linear least-squares fits to the numerical values, respectively. Least-squares fitting is done neglecting the values computed on the two coarsest grids.}
\label{Fig:oaccuracy}
\end{figure}

Figure~\ref{Fig:d2species}a shows the numerical and analytical solutions of time history of $\left({d}/{d_0}\right)^2$ in the stagnant conditions for the mass transfer numbers of $B_M = 0.025, 0.05$ and $0.1$. Simulations are repeated using $128\times 256$, $256\times 512$ and $512\times 1024$ grid resolutions to demonstrate the grid convergence. 
As seen, the numerical results are in good agreement with the analytical solution and the $128\times 256$ grid resolution is sufficient for the grid convergence. Simulations are also performed to examine effects of the domain size and the results are plotted in Fig.~\ref{Fig:d2species}b. 
This figure shows that the numerical results converge to the analytical solution as the domain size increases as expected since the $d^2$-law assumes that the droplet evaporates in an unbounded domain. However, increasing the domain size from $5d_0 \times 10d_0$ to $10d_0 \times 20d_0$  does not a significant effect on the overall results for this case. 
The spatial accuracy is quantified in Fig.~\ref{Fig:oaccuracy}. As seen in Fig.~\ref{Fig:oaccuracy}a, the four finest grids are in the asymptotic range and the nearly linear relationship between $(d/d_0)^2$ and $(h/d_0)^2$ indicates a second order spatial accuracy, which is verified more rigorously in Fig.~\ref{Fig:oaccuracy}b.  Following \citet{Muradoglu-etal-cf-2003}, the error-free values of $\left({d}/{d_0}\right)^2$ used to compute the spatial error in Fig.~\ref{Fig:oaccuracy}b are estimated using the Richardson's extrapolation based on the least-squares fits as $h\rightarrow 0$. The error is then computed as the absolute value of difference between the numerical solution and the predicted spatial error-free value. The linear least-squares fits show that the present numerical method is second order accurate in space. Note that the spatial accuracy was found to be 1.3 for the same case in the previous studies of \citet{irfan2017front,irfan2018front} where the mass source term was conservatively distributed outside of the interface using the one-sided adsorption layer method \citep{muradoglu2008front,MURADOGLU2014737}. The results demonstrate that the present method significantly improves the spatial accuracy compared to the one-sided distribution of the mass source term as done in \citep{irfan2017front,irfan2018front}. 

\subsection{Wet-bulb temperature comparison}
In this test case, the wet-bulb temperature of a water droplet is computed at various ambient conditions and compared to that of the psychometric chart to demonstrate a proper coupling of the species and temperature fields. A water droplet evaporates solely due to the vapor mass fraction gradient at the interface. As the evaporation proceeds, the temperature at the interface reduces and eventually reaches a steady value called the wet-bulb temperature that is uniquely determined by the relative humidity (RH) and dry-bulb temperature of the ambient air. 

\begin{figure}[ht]
\centering
\includegraphics[scale=0.415]{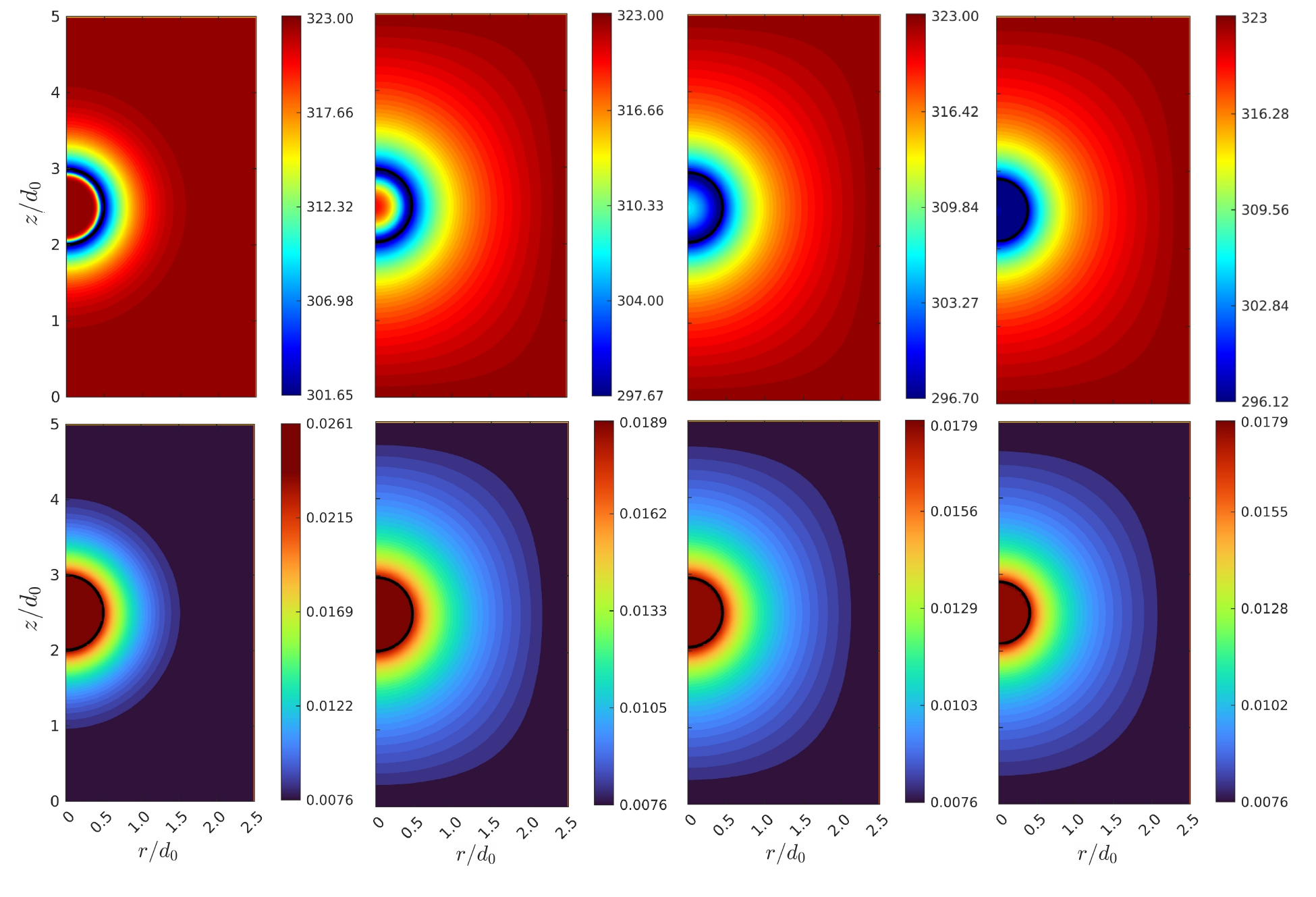}
\caption{Evolution of temperature (top row) and vapor mass fraction (bottom row) fields for a water droplet evaporating at ambient dry-bulb temperature of 323.15 K and relative humidity of 10\% at the non-dimensional times (from left to right)  $t^*= 0.02,\; 0.20, \; 2.80,$ and $7.91$. (Domain: $2.5d\times 5d$;  Grid: $128\times 256$)}
\label{Fig:wetbulb 323}
\end{figure}

\begin{figure}[ht]
\centering
\begin{tabular}{cc}
\includegraphics[scale=0.3]{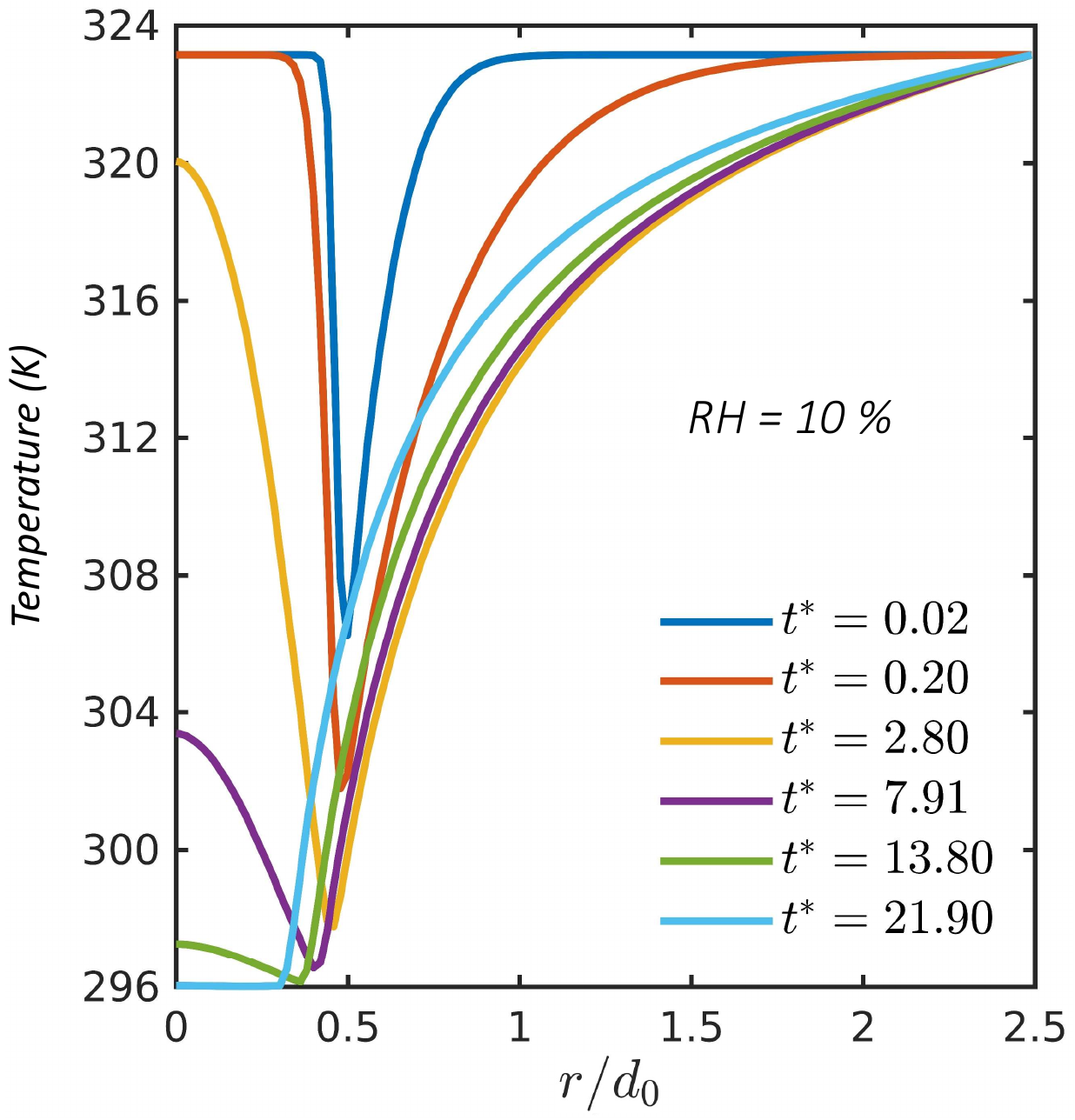}
&
\includegraphics[scale=0.3]{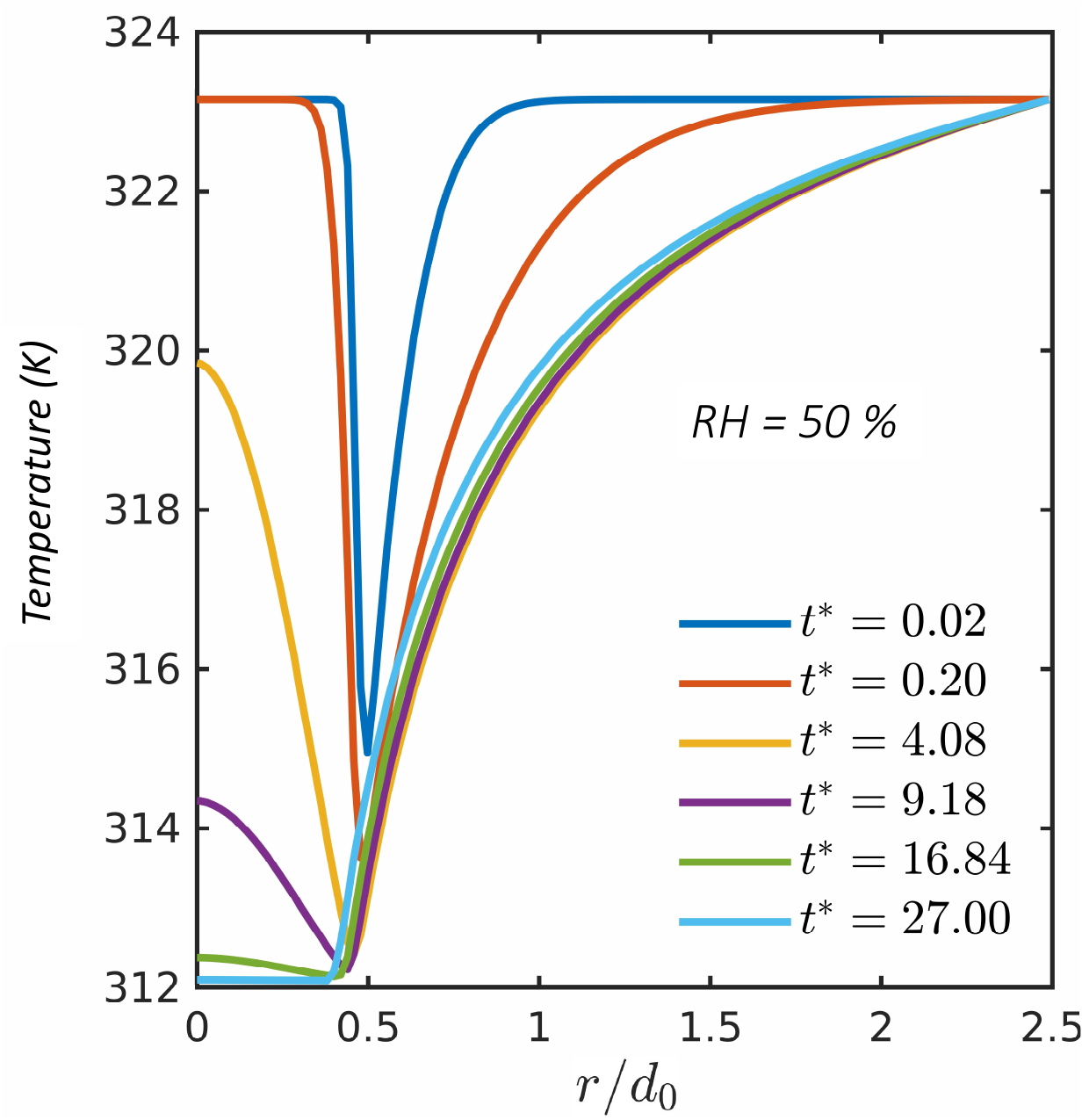}\\
\small{(a)} & \small{(b)}
\end{tabular}
\caption{ Temperature profiles plotted along the centerline of the droplet in the radial direction at various non-dimensional times. Dry bulb temperature is 323.15 K. Relative humidity is  (a) 10\% and (b) 50\%. (Domain: $2.5d\times 5d$;  Grid: $128\times 256$) }
\label{Fig:tempevol}
\end{figure}

\begin{figure}[ht]
\centering
\begin{tabular}{cc}
\includegraphics[scale=0.68]{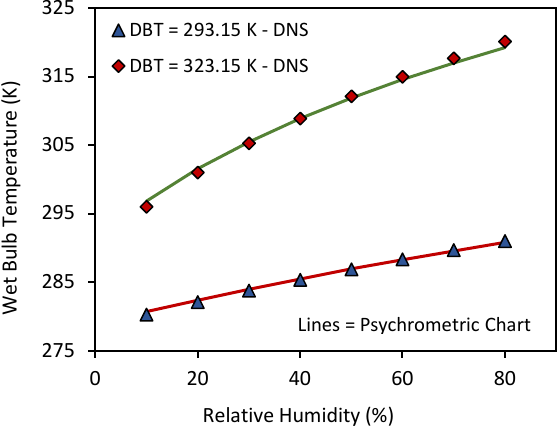}
&
\includegraphics[scale=0.68]{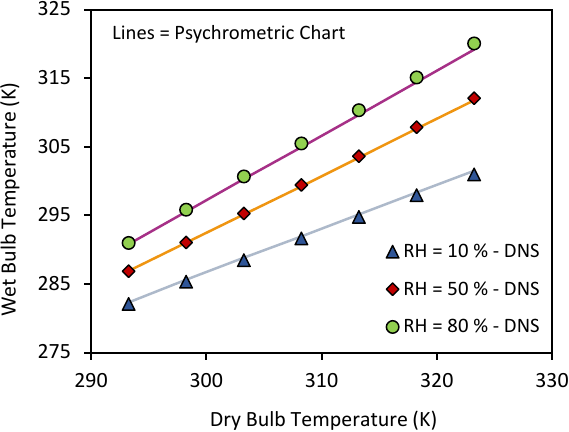}\\
\small{(a)} & \small{(b)}
\end{tabular}
\caption{(a) Variation of the wet-bulb temperature with respect to the relative humidity at fixed dry-bulb temperatures of 293.15 K and 323.15 K. (b) Effect of increasing dry-bulb temperature at a fixed relative humidity of 10\%, 50\%, and 80\%. (Domain: $2.5d\times 5d$;  Grid: $128\times 256$)}
\label{Fig:wetbulbplots}
\end{figure}

In the simulations, a water droplet of initial diameter $d_0$ is placed centered at ($0,2.5d_0$) in a computational domain of $2.5d_0\times 5d_0$ that is resolved by a $128\times 256$ uniform Cartesian grid.  The temperature is initially set to a constant dry-bulb temperature $T_g$ in the entire computational domain. The temperature and vapor mass fraction are fixed at $T_g$ and $Y_{vap}$ as Dirichlet boundary conditions at the domain boundaries where $Y_{vap}$ is computed as $Y_{vap} = \omega_h/(1+\omega_h)$ with $\omega_h$ being the humidity ratio read from a psychrometric chart as a function of dry-bulb temperature and relative humidity. The actual material properties of water and air are used in the simulations except for the water density which is set to $\rho_l= 5\rho_g$. The thermal conductivity of water, $k_l$, is adjusted to match the physical thermal diffusivity~\citep{irfan2017front}. The results are non-dimensionalized using $d_0$ and $d^2_0/D_{vg}$ as the length and time scales, respectively. Figure~\ref{Fig:wetbulb 323} shows the evolution of the temperature and species fields at various instances for the dry-bulb temperature of $T_g=323.15$~K and relative humidity of $RH=10\%$. As the droplet evaporates, the interface temperature decreases and vapor diffuses deeper into the ambient air. This thermal exchange continues until the heat absorbed by the evaporation exactly balances the heat conducted from the ambient air and the water droplet reaches the wet-bulb temperature. Figure~\ref{Fig:tempevol} shows the temporal evolution of radial temperature profiles along the horizontal center line of the domain width for the cases of $RH = 10\%$ and $RH = 50\%$. As seen, cooling starts at the interface ($r/d_0=0.5$) and gradually diffuses into the droplet until the droplet reaches a constant wet-bulb temperature. The computed wet-bulb temperature is compared with the values read from the psychometric chart in Fig.~\ref{Fig:wetbulbplots}.  The wet-bulb temperature is first plotted in Fig.~\ref{Fig:wetbulbplots}a against the relative humidity in the range $10\%\le RH\le 80\%$ for the dry-bulb temperature values of 293.15~K and 323.16~K, and then against the dry-bulb temperature in the range $293.1~{\rm K}\le T_g \le 323.15~{\rm}K$ for the fixed relative humidity values of $10\%$, $50\%$ and $80\%$ in Fig.~\ref{Fig:wetbulbplots}b. As seen, the numerical results are in good agreement with the psychometric chart values in both cases demonstrating accurate coupling of the temperature and vapor mass fraction fields. 

\subsection{Droplet Evaporation in a Convective Environment}
\label{EvapConv}

It is computationally not feasible to use the interface-resolved techniques in large scale simulations of evaporating sprays and droplet clouds relevant to industrial applications and environmental flows. Instead, the Lagrangian point particle methods are widely used in modeling the evaporation of sub-grid droplets in such applications \cite{JENNY2012846}. In this approach, the droplets are modeled as Lagrangian point-particles and evaporation models are used to compute the energy and mass transfer rates between droplets and the ambient fluid. Success of such simulations is critically dependent on the accuracy of the low-order evaporation models. A detailed review of the low-order models can be found in the review papers by \citet{FAETH1977191}, \citet{SIRIGNANO1983291}, \citet{LAW1982171} and \citet{sazhin2006advanced,SAZHIN201769}. 

The interface-resolved simulations can provide indispensable insight into the complex dynamics of droplet evaporation needed for the assessment and improvement/development of the low-order models. Previous studies  \citep{SCHLOTTKE20085215, ReutzschEtAl-jcp-2020,HAYWOOD19941401} 
 have predominantly focused on reporting the surface-averaged Sherwood number for a droplet evaporating in a convective environment. However, information regarding  distribution of the Sherwood number at interface, particularly for a deforming droplet, is very scarce. Employing the IB/FT method, we demonstrate how droplet deformation can fundamentally alter the distribution of Sherwood number. 
In this section, the present numerical method is applied to simulate a droplet evaporating in a convective environment and the results are compared with the analytical models for the mass transfer number in the range $1 \leq B_M \leq 15$.

To mimic the evaporation of a droplet in a convective environment, the center of mass of the droplet is fixed in the space using a moving reference frame (MRF) methodology, and the momentum equation is modified accordingly to account for the acceleration of the reference frame \citep{rusche2003computational}. In the experimental studies, the droplet is usually deposited on a fiber to keep it fixed in space in a flowing ambient fluid \citep{chauveau2011analysis,gokalp1989observations,gokalp1992mass,gokalp1994vaporization}. However, in the real-world applications such as spray evaporation and cloud formation, droplets can freely translate  and deform in the ambient fluid flow. Therefore, employing a moving reference frame provides a more relevant framework to study droplet evaporation in a convective environment. Accounting for the acceleration of the MRF, the modified momentum equation can be written in the non-conservative form as \citep{rusche2003computational}
\begin{equation} 
\begin{aligned}
\rho \frac{\partial { \boldsymbol{u}_{rel}}}{\partial {t}} +\rho \left[\nabla \cdot \left(\boldsymbol{u}_{rel}\boldsymbol{u}_{rel}\right) -\boldsymbol{u}_{rel} \left( \nabla \cdot\boldsymbol{u}_{rel}\right) \right]= - \nabla p+\rho \left(\boldsymbol{g} + \boldsymbol{a}_{MRF}\right) \\ + \nabla \cdot \mu \left(\nabla \boldsymbol{u}_{rel}+\nabla {\boldsymbol{u}}^T_{rel}\right)+ \int_A{\sigma \kappa \boldsymbol{n}\delta \left(\boldsymbol{x}\mathrm{-}{\boldsymbol{x}}_{\mathit{\Gamma}}\right)dA},
\end{aligned}
\label{eqn:momentumrel}
\end{equation}
where $\boldsymbol{u}_{rel}$ is the relative velocity and $\boldsymbol{a}_{MRF}$ is the acceleration needed to hold the centroid of the droplet fixed with respect to the computational domain. In the present study, the procedure proposed by \citet{rusche2003computational} is used to compute $\boldsymbol{a}_{MRF}$. 

\begin{figure}[ht]
\centering
\includegraphics[scale=0.8]{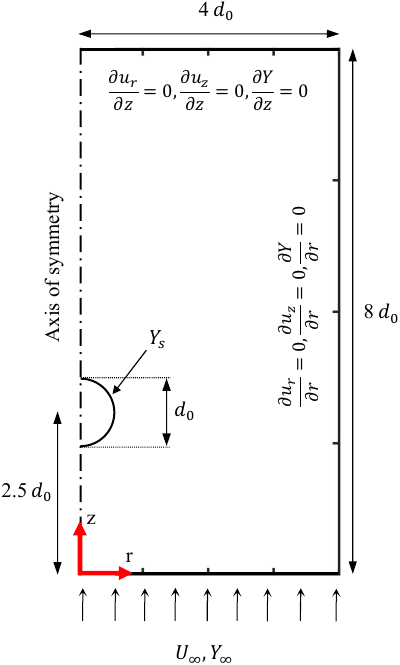}
\caption{Sketch of the computational domain and the boundary conditions used for simulation of droplet evaporation in a convective environment.}
\label{Fig:Domain}
\end{figure}

We consider a deformable droplet whose center is fixed in the space in a flowing ambient fluid as sketched in Fig.~\ref{Fig:Domain}. The ambient flow is uniform far from the droplet and its velocity is $U_{\infty}$. The droplet is initially spherical with a diameter of $d_0$ and its center is located at the axial distance of $2.5d_0$ from the inlet. The computational domain is $4d_0\times 8d_0$ and it is resolved by a $512\times 1024$ uniform Cartesian grid in the radial and axial directions, respectively. In all the simulations, a uniform velocity $U_{\infty}$ is specified at the inlet section corresponding to the specified Reynolds number while the symmetry and full-slip boundary conditions are applied at the left (centerline) and the right (far-field) boundaries. The Dirichlet mass fraction boundary condition, corresponding to a specific mass transfer number ($B_M$), is applied on the droplet surface and the vapor mass fraction is set to zero at the inlet, i.e., the gas is dry.
The gravitational effects are neglected to facilitate a direct comparison with the evaporation models.  

The relevant non-dimensional parameters for this study are defined as
\begin{eqnarray}
Re=\frac{{ \rho }_gU_{\infty}d_0}{{\mu }_g},  \; We= \frac{\rho_g {U_{\infty}}^2 d_0}{\sigma}, \; Sc=\frac{{\mu }_g}{{\rho }_gD_{vg}}, 
\label{eqn:nodimensional}
\end{eqnarray}
where $Re$, $We$ and $Sc$ are the Reynolds, Weber and Schmidt numbers, respectively. The Reynolds and Schmidt numbers are fixed at $Re=100$ and $Sc = 0.7$ in all the results presented here. Simulations are carried out for three different Weber numbers representing a nearly spherical ($We= 0.65$), a moderately deformable ($We = 6.5$) and a highly deformable ($We=13$) cases. The viscosity and density ratios constitute the other relevant non-dimensional numbers and they are fixed at $\mu_l/\mu_g=15.34,$ and $\rho_l/\rho_g=25.75$, respectively. Note that the subscripts $'l'$ and $'g'$ denote the properties of the droplet liquid and the ambient gas, respectively. 

\begin{figure}[ht]
\centering
\includegraphics[scale=0.4]{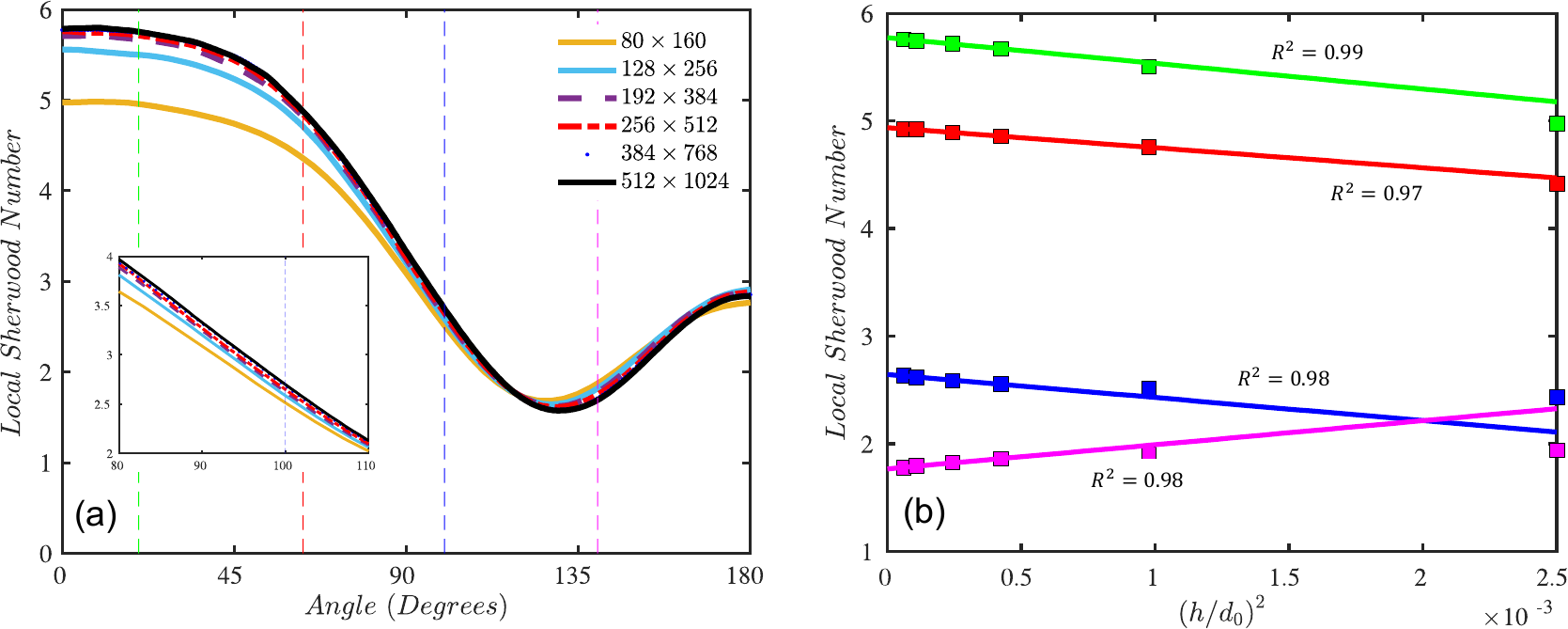} \\

\includegraphics[scale=0.4]{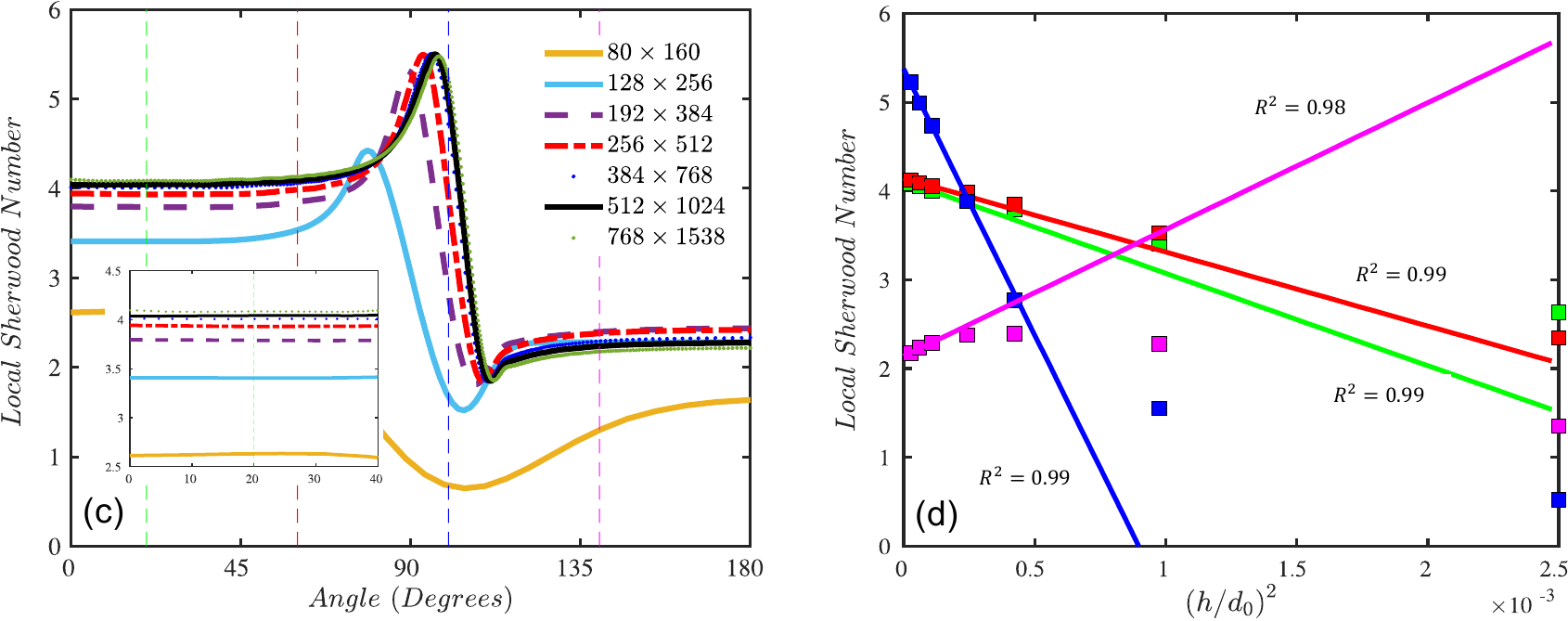}

\caption{Local Sherwood number computed over (a) nearly spherical ($We=0.65$) and (c) highly deformable ($We=13$) droplets for $Re=100$ and $B_M = 2$ at $t^*=5$. Grid resolutions: $80 \times 160, 128 \times 256, 192 \times 384, 256 \times 512, 384 \times 768,$ $512 \times 1024,$ and $768 \times 1536$.  Quantification of the spatial accuracy for (b) the nearly spherical ($We=0.65$) and (d) the highly deformable ($We=13$) droplets. }
\label{Fig:evap-ordac}
\end{figure}

Simulations are first performed to demonstrate grid convergence and to quantify the spatial accuracy of the numerical method for this challenging test case. For this purpose, the effect of grid refinement on the local Sherwood number is considered. Figures~\ref{Fig:evap-ordac}a and ~\ref{Fig:evap-ordac}c show the distribution of local Sherwood number computed at $t^* = 5$ using various grid resolutions in the range between $80 \times 160$ and $768 \times 1536$ for a nearly spherical ($We = 0.65$) and a highly deformable ($We = 13$) cases. As seen, the difference between the results obtained on successive grid resolutions decreases with grid refinement, indicating a grid convergence. The spatial accuracy is quantified in Figs.~\ref{Fig:evap-ordac}b and ~\ref{Fig:evap-ordac}d where the local Sherwood number is plotted against the square of the normalized grid size $\left(h/d_0\right)^2$ at various angles indicated by the dotted vertical lines in Figs.~\ref{Fig:evap-ordac}a and ~\ref{Fig:evap-ordac}c. In these figures, symbols denote numerical values while solid lines are linear least-squares fits to the numerical data. Note that only the numerical values obtained on the grid resolutions in the asymptotic range are used in the linear least-squares fits. The nearly linear relationship between the numerical values and linear least-squares fits indicates that the numerical method is second order accurate in space, which constitutes a major advantage of the present hybrid method compared to the method developed by \citet{irfan2017front,irfan2018front} where the spatial accuracy was only first order. Figure~\ref{Fig:evap-ordac} shows that the $512\times 1024$ grid resolution is sufficient to reduce the spatial error below 0.5\% and 7\% for nearly spherical ($We = 0.65$) and highly deformable ($We = 13$) cases, respectively. Therefore, this grid resolutions is used in all the subsequent simulations. The larger numerical  error for the most deformable case is due to a significantly fewer number of grid points inside the deformed droplet compared to that in the nearly spherical one. For instance, number of grid points along the axis of symmetry inside the droplet is 130 and 40 for the nearly spherical ($We = 0.65$) and the highly deformed ($We = 13$) cases, respectively, i.e., the nearly spherical droplet is about three times better resolved. 

Figure~\ref{Fig:evap-ordac} shows that the variation of the
local Sherwood number over the deformed droplet ($We = 13$) is fundamentally different than that of the nearly spherical one ($We = 0.65$). For instance, the local Sherwood number becomes maximum at the centerline on the leading edge for the nearly spherical droplet while it occurs in the shoulder region of the deformed droplet. This marked difference clearly demonstrates the importance of droplet deformation that is totally ignored in the commonly used low-order evaporation models.

\begin{figure}[ht]
\centering
\begin{tabular}{cc}
\includegraphics[scale=0.54]{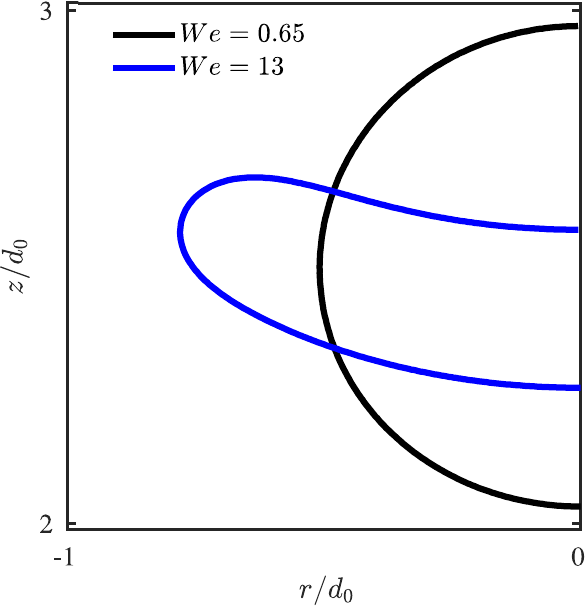} 
&
\includegraphics[scale=0.5]{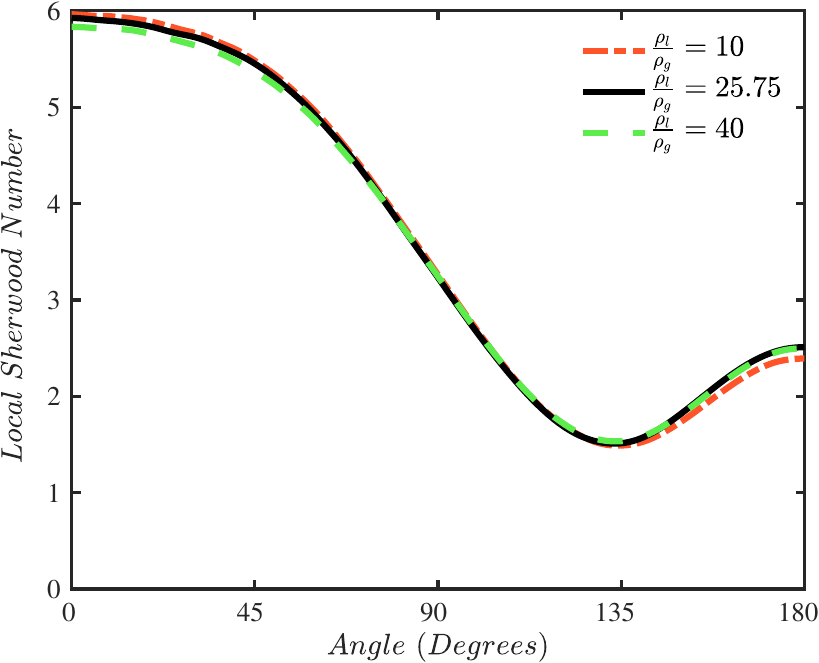} \\
\small{(a)} & \small{(b)}
\end{tabular}
\caption{(a) Droplet shapes for the nearly spherical ($We = 0.65$) and highly deformable ($We = 13$) cases at $t^* = 5$. (b) Effect of density ratio  on Sherwood number distribution for the $We=0.65$ case. ($Re=100$, $Sc = 0.7$, $B_M = 2$. Grid: $512 \times 1024)$. }
\label{Fig:rhoratio}
\end{figure}

The density ratio is of order of $1000$ in actual spray combustion simulations but the well-known inconsistency in the advection schemes used to advect the density and the momentum causes numerical difficulties in the present front-tracking method at very high density ratios \citep{tryggvason2011direct}. Therefore the density ratio is usually limited to $\rho_l/\rho_g \le 40$. \citet{OlgacEtAl_cf_13} and \citet{TasogluEtAl_pof_10} have shown that the results are not affected significantly when the density ratio is further increased beyond $\rho_l/\rho_g = 20$. Simulations are performed for the range of density ratio between $10\le \rho_l/\rho_g \le 40$ to examine  sensitivity of  results to density ratio and to determine the density ratio beyond which the results are not affected significantly. The other parameters are fixed at $Re=100$, $Sc = 0.7$, $We=0.65$, $B_M=2$ and  $ \mu_l/\mu_g=15.34$. The results are shown in Fig.~\ref{Fig:rhoratio}b. As seen, the results are insensitive to variation in the density ratio beyond $\rho_l/\rho_g\ge 25.75$. Thus, the density ratio is kept constant at  $\rho_l/\rho_g= 25.75$ in all the simulations presented in this section unless specified otherwise.

\begin{figure}[ht]
\centering
\begin{tabular}{ccc}
\includegraphics[scale=0.46]{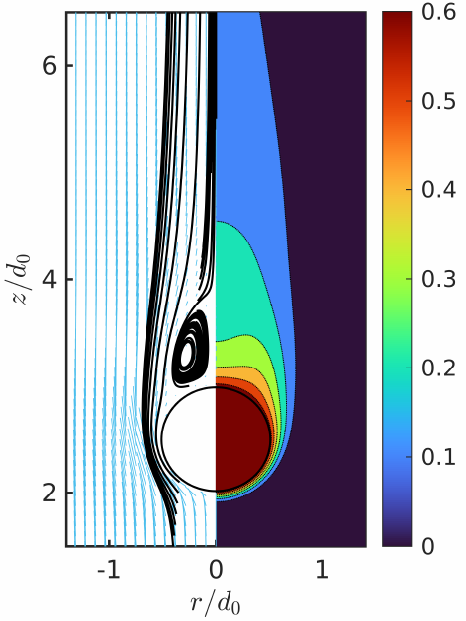} 
&
\includegraphics[scale=0.46]{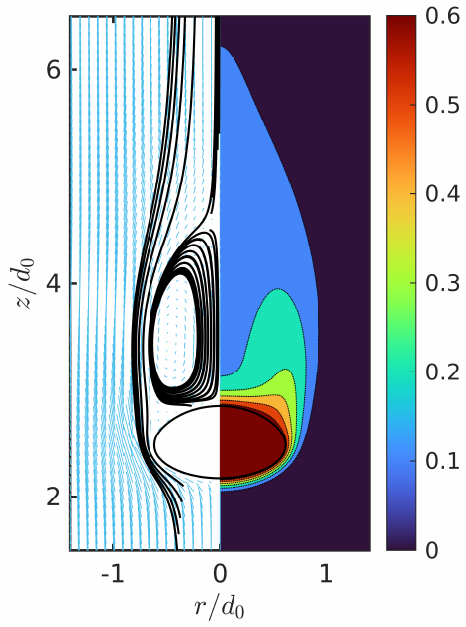} 
&
\includegraphics[scale=0.46]{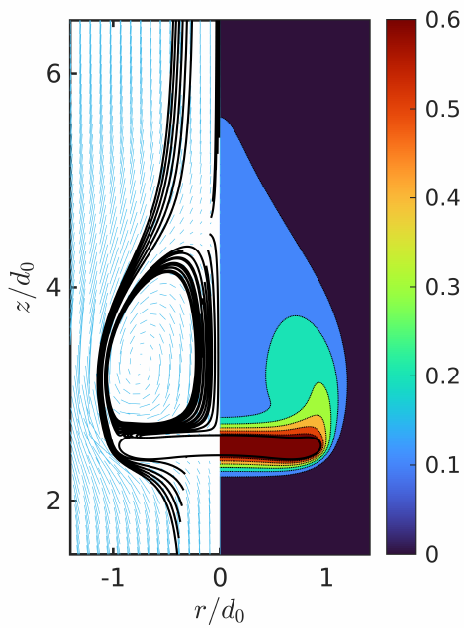} \\
\includegraphics[scale=0.46]{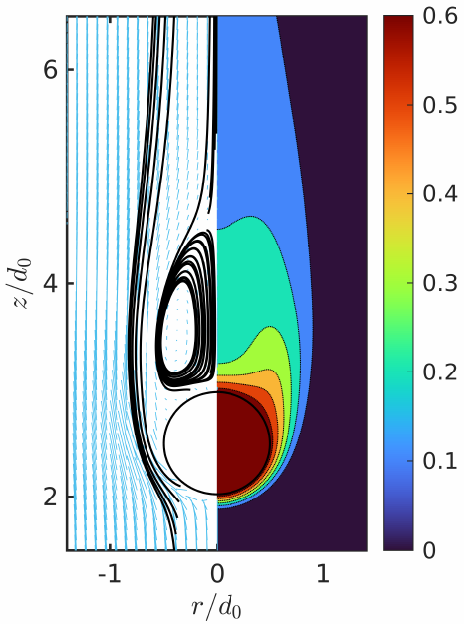} 
&
\includegraphics[scale=0.46]{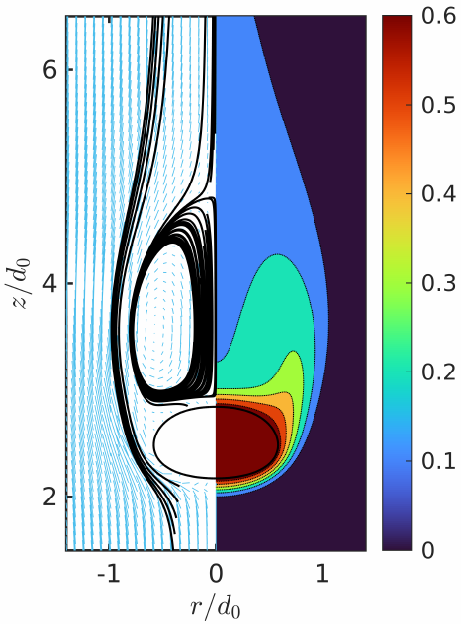} 
&
\includegraphics[scale=0.46]{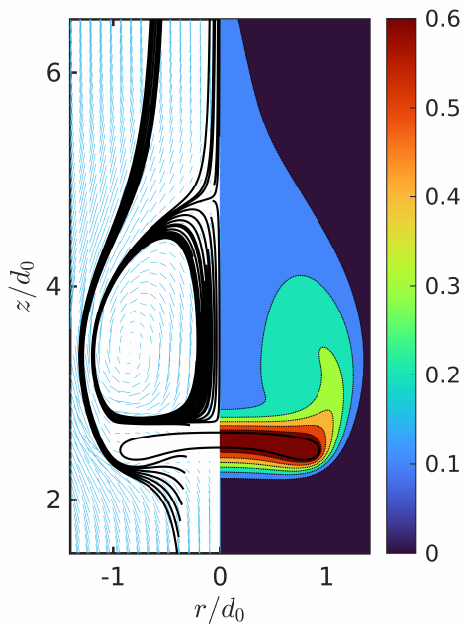} \\

\small{(a)} & \small{(b)} & \small{(c)}
\end{tabular}
\caption{Effects of droplet deformability. Velocity vectors and streamlines (left portion) and vapor mass-fraction field (right portion) are shown for the non-evaporating (no Stefan) (top row) and evaporating ($B_M=2$) (bottom row) cases at $t^*=9$. Weber numbers are 0.65 (a), 6.5 (b), and 13 (c) from left to right. The other parameters are $Re=100$, $\rho_l/\rho_g =25.75$, $\mu_l/\mu_g=15.34$ and $Sc = 0.7$.  Color bars indicate the values of mass fraction. (Domain: $4d_0\times 8d_0$; Grid:$512\times 1024$). }
\label{Fig:We06}
\end{figure}

After demonstrating grid convergence and insensitivity to the density ratio, the numerical method is applied to simulate non-evaporating and evaporating droplets for $We= 0.65$, 6.5, and 13 to show its capability in resolving evaporation of nearly spherical and highly deformable droplets in a convective environment. It is emphasized here that the term of ``{\it non-evaporating droplet}" is used here to refer to the case that the mass transfer from the droplet does not induce any Stefan flow, i.e., the Stefan flow is switched off manually to demonstrate its sole effect on the flow field and the mass transfer from the droplet. Figure~\ref{Fig:We06} illustrates the velocity vectors, streamlines and mass fraction field around non-evaporating (i.e., no Stefan flow) and evaporating ($B_M=2$) droplets at $Re=100$. As seen, a flow separation occurs in all the cases creating a recirculation zone behind the droplet that broadens as the droplet deformation (i.e., $We$) increases \cite{SETIYA2023104455}. In the evaporating cases, the streamlines are pushed away from the droplet surface due to the Stefan flow, resulting in a thickened boundary layer that leads to an early flow separation and a broader recirculation zone compared to the corresponding non-evaporating cases. In addition, the front and back stagnation points are detached from the surface of the droplet for the same reason in the evaporating cases. As can be seen in Figs.~\ref{Fig:We06}(b) and ~\ref{Fig:We06}(c), the Stefan flow also slightly influences deformation of droplet.

\begin{figure}[!t]
\centering
\begin{tabular}{ccc}
\includegraphics[scale=0.53]{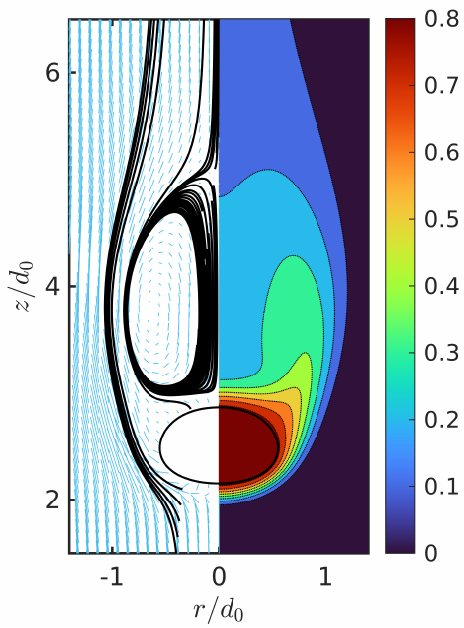} 
&
\includegraphics[scale=0.53]{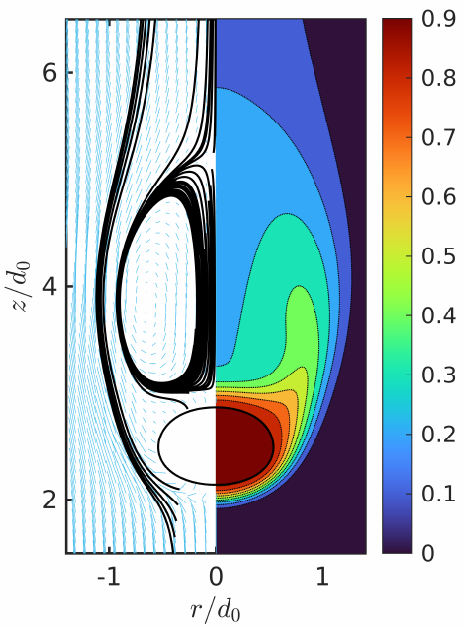} 
&
\includegraphics[scale=0.53]{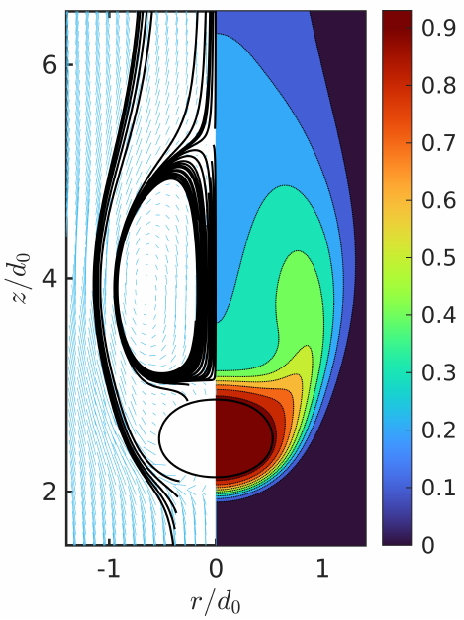} \\
\small{(a)} & \small{(b)} & \small{(c)}
\end{tabular}
\caption{Effects of evaporation intensity. Velocity vectors and streamlines (left portion) and vapor mass-fraction field (right portion) around a moderately deformable evaporating droplet at $Re=100$, $t^*=9$, and mass transfer numbers of (a) $B_M = $5, (b) 10, and (c) 15.  Color bars indicate the value of vapor mass-fraction. (Domain: $4d_0\times 8d_0$; Grid:$512\times1024$). }
\label{Fig:highBM}
\end{figure}

To better show the effects of the Stefan flow, further simulations are performed for the moderately deforming droplet case of $We = 6.5$ at various evaporation intensities, i.e., at $B_M = 5$, 10 and 15, which can be relevant in high temperature spray combustion applications.  The results are plotted in Fig.~\ref{Fig:highBM}. As the Stefan flow intensifies with increasing $B_M$, the streamlines are pushed further away from the droplet interface and the recirculation zone gets enlarged. This figure clearly shows the importance of the Stefan flow which can be as large as that of the mean flow and modify the entire flow field around the droplet. These results also indicate that the present numerical method can successfully simulate the evaporating droplets with high evaporation intensities. 

Finally, we compare the interface-resolved simulation results with the commonly used low-order evaporation models, i.e., the classical~\citep{sazhin2006advanced} and Abramzon and Sirignano (A-S)~\citep{abramzon1989droplet, sirignano2010fluid} models. Both models ignore droplet deformation and assume that droplet remain spherical during the entire evaporation process. The classical model takes into account effect of the Stefan flow on mass transfer rate from evaporating droplets ~\citep{sazhin2006advanced} but ignores the boundary layer thickening caused by the surface blowing. According to the classical model, the Sherwood number is given by
\begin{equation} 
\begin{aligned}
 Sh= Sh_0 \frac{\ln\left(1+B_M \right)}{B_M },
\end{aligned}
\label{eqn:Classicalmodel}
\end{equation}
where $Sh_0$ is the Sherwood number for the non-vaporizing spherical particle and it is usually estimated using the correlations obtained for the mass transfer from a solid sphere. For instance, the Frossling correlation~\citep{frossling1938evaporation} is widely used in spray simulations and it is given by
\begin{equation} 
\begin{aligned}
Sh_0 = 2 + 0.552 \ {Re}^{\frac{1}{2}} \ {Sc}^{\frac{1}{3}}.
\end{aligned}
\label{eqn:Frossling}
\end{equation}

Abramzon and Sirignano \citep{abramzon1989droplet, sirignano2010fluid} proposed a modification to the classical model to account for the effects of boundary layer thickening due to the surface blowing of the Stefan flow. They took into account the adverse effect of Stefan flow on the Sherwood number by approximating the droplet as a collection of evaporating wedges. They found that the correction factor is essentially dependent on the mass transfer number, and proposed the following correction to quantify its effect on the Sherwood (or Nusselt) number
\begin{equation}
\begin{aligned} 
   & Sh = 2+\frac{Sh_0-2}{F_M}, \\
   & F_M = {\left(1+B_M \right)}^{0.7} \ \frac{\ln\left(1+B_M \right)}{B_M} 
\end{aligned}
\label{eqn:A-S}
\end{equation}

Note that the Frossling correlation (i.e., Eq.~\eqref{eqn:Frossling}) gives an average value of the Sherwood number for a spherical particle but it does not provide any information for the local mass transfer rate. Thus, the local Sherwood number, i.e., the distribution of $Sh_0$, is directly obtained from the sharp-interface immersed boundary method simulations, and it is used in evaluating the classical and  Abramzon and Sirignano models in the present study. 

\begin{figure}[!t]
\centering
\includegraphics[scale=0.5]{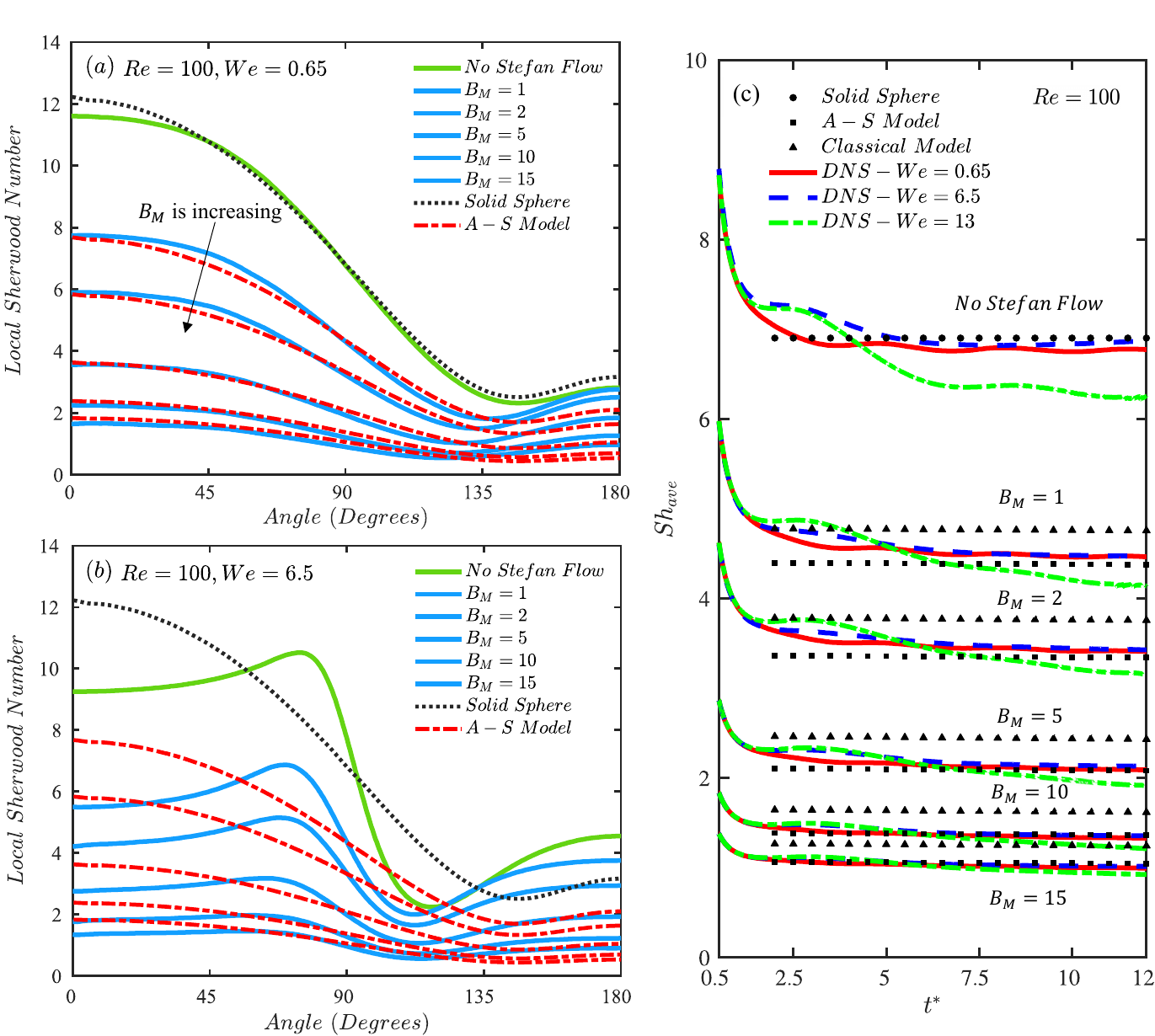}
\caption{Interface-resolved simulation results for the local Sherwood number over (a) a nearly spherical and (b) a moderately deformable droplet in comparison with the Abramzon-Sirignano model ($Re=100$ and $t^*=9$). (c) Temporal evolution of the interface-resolved surface-averaged Sherwood number for a nearly spherical $\left(We=0.65\right)$, a moderately deformable $\left(We=6.5\right)$, and a highly deformable $\left(We=13\right)$  droplet at $Re=100$. The numerical results are compared with the classical and Abramzon-Sirignano models. The Sherwood number computed for a solid sphere is also shown on the right. }
\label{Fig:ShRe100We6andWe06}
\end{figure}

Figures~\ref{Fig:ShRe100We6andWe06}a and ~\ref{Fig:ShRe100We6andWe06}b, respectively, show the variation of the local Sherwood number over a nearly spherical and a moderately deforming evaporating droplet for the range of mass transfer numbers between $1 \leq B_M \leq 15$ at $t^*=9$. The numerical results are compared with the Abramzon-Sirignano model where the local $Sh_0$ is taken from separate the sharp-interface immersed-boundary simulations performed for the mass transfer from a solid sphere of the initial droplet radius. Note that the change in droplet diameter is neglected in evaluating $Sh_0$ since the evaporative mass loss causes less than 5\% decrease in droplet diameter until $t^* \le 9$, resulting in less than 1.75\% reduction in the surface-averaged Sherwood number. In addition, as mentioned before, the Stefan flow is switched off in the case of {\it No Stefan Flow} plotted in Fig.~\ref{Fig:ShRe100We6andWe06} to demonstrate its sole effect. As seen, the Abramzon-Sirignano model agrees reasonably well with the numerical results for the nearly spherical droplet case especially in the leading edge until a flow separation occurs. In the wake region, it generally underpredicts the mass transfer rate. As droplet undergoes a significant deformation, its performance deteriorates quickly resulting in a qualitatively inaccurate prediction of local Sherwood number all over the droplet especially at smaller mass transfer numbers. 

Figure~\ref{Fig:ShRe100We6andWe06}c illustrates the temporal variation of the surface-averaged droplet Sherwood number computed for three different Weber numbers of $We= 0.65, 6.5,$ and $ 13$, in comparison with the Abramzon-Sirignano and the classical models. Note that, in this figure, $Sh_0$ is evaluated using the Frossling correlation based on the instantaneous Reynolds number of the moderately deforming droplet ($We=6.5$) for each mass transfer number. The equivalent droplet diameter is used in evaluating the Sherwood number for deformable cases. We also note that the maximum reduction in the droplet diameter is about 7.3\% at the end of the simulation for the highly deformable case at $B_M=15$ in Fig.~\ref{Fig:ShRe100We6andWe06}c. According to the Frossling correlation, this reduction results in a maximum decrease of 2.6\% in the Sherwood number. Figure~\ref{Fig:ShRe100We6andWe06}c shows that the Abramzon-Sirignano model outperforms the classical model for all the mass transfer numbers. As seen, the Abramzon-Sirignano model slightly under-predicts while the classical model considerably over-predicts the average Sherwood number. The discrepancy between the Abramzon-Sirignano and DNS is consistent with the analysis shown in Figs.~\ref{Fig:ShRe100We6andWe06}a and \ref{Fig:ShRe100We6andWe06}b.

Figure \ref{Fig:ShRe100We6andWe06}c also reveals the transient effect of droplet deformability on the surface-averaged Sherwood number. Note that $Sh_{ave}$ remains about the same for all $B_M$ values regardless of the droplet Weber number until the droplet undergoes a significant deformation, i.e., until $t^*=1.2$. Then, the effects of droplet deformability becomes visible in the surface-averaged Sherwood number. It is interesting to observe that the surface-averaged Sherwood number becomes very similar for the nearly spherical and moderately deformed cases when a quasi steady-state is attained. For instance, at $B_M=1$, the averaged Sherwood number is found to be $Sh_{ave}=6.77$ and $Sh_{ave}=6.89$ for the nearly spherical and the moderately deformed  cases, respectively. The time evolution of $Sh_{ave}$ is also similar for these cases. However, in the highly deformed droplet case, $Sh_{ave}$ deviates considerably and the difference becomes as large as 8\%. More importantly, the time evolution is also qualitatively different in this case. This figure also shows that  the Abramzon-Sirignano model is very successful in predicting the average Sherwood number especially in the nearly spherical case. It is interesting to observe that, in spite of the qualitative differences in the local Sherwood number distributions as shown in Fig.~\ref{Fig:ShRe100We6andWe06}b, there is much better agreement between the fully-resolved results and the Abramzon-Sirignano model predictions for the average Sherwood numbers. The increased surface area of the deformed droplet leads to an overall higher mass transfer rate as also reported by \citet{SETIYA2023104455}. However, despite this over all increase, the surface-averaged Sherwood number appears to be lower for the highly deformed case as demonstrated in Fig. \ref{Fig:ShRe100We6andWe06}b.

\section{Conlusions}
\label{conlusions} 
This study presents a novel hybrid sharp-interface immersed-boundary/front-tracking method for interface-resolved simulations of a droplet evaporation in incompressible multiphase flows. A distinctive feature of this work is the incorporation of an image point and ghost cell methodology into the front-tracking framework to achieve a second order spatial accuracy for the mass transfer at the interface. The methodology is also used successfully to impose the no-slip boundary conditions on a solid sphere.

The multiphase flow solver is validated for a falling droplet case. Then the numerical method is applied to simulate a single phase flow over a solid sphere in a uniform ambient flow to show the accuracy of the sharp-interface immersed-boundary method. The results are found to be in excellent agreement with the published numerical results for the both cases. 

The evaporation model is first validated for the classical $d^2$-law test case and the numerical results are found to be in excellent agreement with the analytical solution. It is demonstrated that the method is grid-convergent and second-order accurate in space, surpassing the previous front-tracking method that employs an absorption layer to impose the mass transfer boundary conditions indirectly and results in only a first order spatial accuracy~\citep{irfan2017front,irfan2018front}. The method is then applied to simulate the wet-bulb temperature of a water droplet under various ambient conditions to show its performance for two-way coupling of species and energy fields, and the numerical results are found to be in good agreement with the psychrometric chart values.  

The method is finally applied to study droplet evaporation in a convective environment and the results are compared with the classical and Abramzon-Sirignano evaporation models. Using a moving reference frame methodology, simulations are performed for a nearly spherical, a moderately deformable and a highly deformable droplet cases to show the effects of droplet deformation on evaporation. The Reynolds number is kept constant at $Re =100$ but the  the mass transfer number is varied in the range of $1	\leq B_M \leq 15$ to show the ability of the numerical method to simulate slowly and rapidly evaporating cases. It is demonstrated that the numerical method is grid convergent and it is second order accurate in space for this more challenging test case. It is found that a flow separation occurs on the shoulder of the droplet and a large recirculation zone is created behind the droplet at this Reynolds number. The Stefan flow thickens the boundary layer and results in an earlier flow separation and a larger recirculation zone compared to the corresponding non-evaporating case. This effect is amplified as the mass transfer number is increased. The numerical results are compared with the classical and Abramzon-Sirignano models in terms of the local and surface-averaged Sherwood numbers. It is found that the the Abramzon-Sirignano model performs well in predicting the Sherwood number on the leading edge of the droplet but its performance deteriorates significantly  in the recirculation zone behind the droplet, i.e., after the separation point. It is generally found that the the low-order models fall short in accurately predicting the local Sherwood number in the wake region. 

The results clearly demonstrate that the present hybrid method is grid convergent with a second order spatial accuracy and it is well suited for interface-resolved simulation of droplet evaporation even in largely deforming and rapidly evaporating conditions in a strong convection relevant to spray combustion applications.


\section*{Acknowledgment}
We acknowledge financial support from the Scientific and Technical Research Council of Turkey (TUBITAK) [Grant Number 121M361].

\section*{CRediT authorship contribution statement}
\textbf{F. Salimezhad:} Conceptualization, methodology, investigation, software, validation, formal analysis,  visualization, writing – original draft.
\textbf{H. Turkeri:} Project administration, funding acquisition, investigation, review \& editing.
\textbf{I. Gokalp:} Funding acquisition, investigation, review \& editing.
\textbf{M. Muradoglu:} Conceptualization, software, resources, investigation, writing – review \& editing, supervision, project administration, funding acquisition.

\section*{declaration of competing interest}
The authors declare that they have no known competing financial interests or personal relationships that could have appeared to influence the work reported in this paper.




  \bibliographystyle{elsarticle-num-names} 
  \bibliography{References}





\end{document}